\documentclass[12pt,epsfig,eps]{article}
\usepackage{epsfig}
\usepackage{graphicx}
\usepackage{amssymb}
\usepackage{amsmath}
\usepackage{epsfig}
\usepackage{graphicx}
\usepackage{amssymb}
\usepackage{amsmath}
\def\permil{\%\raise.10ex\hbox{$_{\scriptstyle 0}$}}

\def\beq{\begin{equation}}
\def\eeq{\end{equation}}
\def\beqn{\begin{eqnarray}}
\def\eeqn{\end{eqnarray}}
\newcommand{\mbf}[1]{\mbox{\boldmath $#1$}}

\newcommand{\bk}{\mbf{k}}
\newcommand{\bq}{\mbf{q}}
\newcommand{\bp}{\mbf{p}}

\newcommand{\bV}{\mbf{V}}
\def\bea{\begin{eqnarray}}
\def\eea{\end{eqnarray}}
\begin{document}
\hfill
\hspace*{\fill}
\begin{minipage}[t]{3cm}
  DESY-06-185\\
  hep-ph/0610303
\end{minipage}
\vspace*{1.cm}
\begin{center}
\begin{huge}
The Pomeranchuk Singularity and 
Vector Boson Reggeization in Electroweak Theory
\\[1cm]
\end{huge}
\begin{large}
\vspace{0.5cm}
J. Bartels$^a$, L. N. Lipatov$^{a \dagger\; b}$, K. Peters$^c$\\[1cm]
$^a$ II. Institut f\"{u}r Theoretische Physik, Universit\"{a}t Hamburg\\
Luruper Chaussee 149, D-22761 Hamburg, Germany\\
$^b$Petersburg Nuclear Physics Institute\\
Gatchina, 188 300 St.Petersburg, Russia\\
$^c$ School of Physics \& Astronomy, University of Manchester,\\
Manchester M13 9PL, UK
\\[1cm]
\end{large}
\end{center}
\vskip15.0pt \centerline{\bf Abstract} \noindent We investigate
the high energy behaviour of vector boson scattering in the
electroweak sector of the standard model. In analogy with the BFKL
analysis in QCD we compute production amplitudes in the
multi-Regge limit and derive, for the vacuum exchange channel, the
integral equation for vector particle scattering. 
We also derive and solve the bootstrap equations for
the isospin-$1$ exchange channel, both for the reggeizing charged
and non-reggeizing neutral vector bosons. \vskip 1cm \hrule
\vskip 1cm \noindent \noindent $^{(\dagger)}$ {\it Marie Curie Chair of 
Excellence.\\
Work supported in part by the grant RFBR-04-02-17513.}
\vfill
\section{Introduction}
One of the topics to be examined in a future high energy electron positron
collider is the scattering of electroweak vector bosons.
Historically it was the high energy behaviour of these scattering
processes which has led to the requirement of introducing a scalar Higgs boson;
a closer look at the unitarity properties puts bounds on the masses of
the Higgs particle. With the possibility of performing, at the linear
collider, precision experiments of electroweak processes, it will be
necessary to consider electroweak higher order corrections; vector
boson scattering is an important class of processes to be studied
with high accuracy.\\ \\
Unitarity properties of vector scattering reactions are most stringent
near the forward direction where cross sections are large. The object of
central interest is the total cross section, i.e. the nature of the
Pomeranchuk singularity. Related to this is the question
whether the fields of the electroweak sector, in particular the
gauge bosons of the broken gauge group, reggeize; the property of
reggeization provides an indication of a possible compositenes.
It is well known that the gauge bosons of nonabelian gauge theories
reggeize in the leading logarithmic approximation (LLA)
~\cite{GST73, Lip76, BFKL2, BL};
this includes both unbroken gauge symmetries (e.g. QCD)
and spontaneously broken models, such as the (pure) $SU(2)$ Higgs model.
On the other hand, the gauge boson of
the abelian theory of QED seems to be elementary (i.e. non-reggeizing),
at least on that level of accuracy which has been investigated so far.
As to the case of the broken $SU(2)\times U(1)$, the charged vector bosons
lie on Regge trajectories, whereas the situation of the neutral sector is
more complicated: several years ago ~\cite{GS79, LS} strong arguments
have been given that there
exists a neutral Regge pole, but neither the photon nor the $Z$ boson lie on
this trajectory.\\ \\
The best way of exploring the vacuum exchange channel and the
reggeization in the electroweak sector
is by following the calculation of the BFKL Pomeron in QCD:
beginning with the production amplitudes in the multi-Regge region
one derives integral equations which, in the vacuum exchange channel,
describe the elastic scattering and the total cross section, and,
in the isospin-one channel, the reggeization of the vector particles.
In this paper we describe such an analysis of the
electroweak sector of the standard model. As our main results, we present the
integral equation for the scattering amplitude for the vacuum exchange
(`electroweak Pomeron'), and we
construct bootstrap equations to investigate the reggeization in both the
charged and neutral vector bosons exchange channels.\\ \\
This paper is organized as follows. In the following section 2 we
define the setup of our calculations, and we collect the lowest order
results of vector-vector scattering. In section 3 we compute, in the Born
approximation, production amplitudes in the multi-Regge limit. Section 4
contains one and two loop results. In section 5 we write down
the integral equations, and we discuss the solutions,
both for the isospin one exchange channel and for the vacuum channel.
In the following section we describe, as an application of our integral 
equations, the two-loop approximation for elastic $WW$ scattering. 
Concluding remarks are contained in a final section.
\section{Setup and Lowest Order Electroweak Amplitudes}

In this section we define the setup of our calculations, and we collect the
results for vector scattering in the Born approximation. Since we will be
interested on the leading-logarithmic approximation (LLA),
we will neglect fermions.
Let us begin with the simple model, in which the Weinberg angle $%
\vartheta _{W}$ is zero. In this case, the $U(1)$ gauge boson, the $B$-boson,
is a free massless particle, and the $W$-bosons are described by the
isovector field $\overrightarrow{W}_{\mu }$ with mass $M$. The $B$ boson
decouples from the $W$ bosons, and we are dealing with a spontaneously broken
$SU(2)$ models. The polarization vectors $e_{\mu }^{\lambda }$ of the $W$ bosons
in the physical gauge are
\begin{equation}
e_{\mu }^{1,2}=e_{\mu \perp }^{1,2}\,,\,\,e_{\mu }^{3}\simeq \frac{k_{0}}{M}%
\delta _{\mu 3}\,+\frac{k_{3}}{M}\delta _{\mu 0}\,,\,\,k_{\mu }\,e_{\mu
}^{\lambda }=0\,,\,\,(e_{\mu }^{\lambda })^{2}=-1\,,
\end{equation}
where $k=(k_{0},k_{3},0,0)$ is the momentum of the vector boson moving along
the third axis. There is also the Higgs particle with the mass $M_{h}$.
We use Sudakov variables:
\begin{eqnarray}
  k=\alpha p_A+\beta p_B +k_{\perp}\,\,,k_{\perp}^2 = -\bk^2 ,
\end{eqnarray}
where $p_A$ and $p_B$ are two light-like vectors along the 3-direction.
In Regge kinematics we have
\beq
s= (p_{A}+p_{B})^{2}= (2E)^{2}\,\gg -t=-(p_{A^{\prime }}-p_{A})^{2}=%
\overrightarrow{q}^{2}\sim M^{2}\,.
\eeq

The
Born amplitude for the high energy scattering $A+B\rightarrow A^{\prime
}+B^{\prime }$ of the $W$-bosons having definite polarizations $\lambda _{r}$
($r=A,B,A^{\prime },B^{\prime }$) is (see Ref. \cite{Lip76})

\beq
A_{AB}^{(0)A^{\prime }B^{\prime }}=2s\,\,g\,a_{\lambda _{A}}\delta _{\lambda
_{A},\lambda _{A^{\prime }}}T_{A^{\prime }A}^{c}\,\frac{1}{t-M^{2}}%
\,g\,a_{\lambda _{B}}\delta _{\lambda _{B},\lambda _{B^{\prime
}}}T_{B^{\prime }B}^{c}\,
\label{2to2born}
\eeq
with
\beq
a_{1,2}=-1\,,\,\,a_{3}=-\frac{1}{2}\,.
\eeq
For the production of a Higgs particle in $W$-boson collisions
the amplitude also has the factorized form
\begin{equation}
A_{AB}^{(0)hB^{\prime }}=2s\,\,g\,a_{3}\delta _{\lambda _{A},3}\,\delta _{cA}\,%
\frac{1}{t-M^{2}}\,g\,a_{\lambda _{B}}\delta _{\lambda
_{B},\lambda _{B^{\prime }}}T_{B^{\prime }B}^{c}\,.
\label{2to2bornhiggs}
\end{equation}
The isospin generators $T_{A^{\prime }A}^{c}\,$\ in the above expressions
belong to the adjoint representation of the $SU(2)$-group: $T_{A^{\prime
}A}^{c}=-i\varepsilon _{cA^{\prime }A}$.

When generalizing, within the leading logarithmic approximation
(LLA) \beq g^{2}\ln \,\frac{s}{M^{2}}\,\sim 1\,,\,\,g^{2}\sim 1\,,
\eeq the Born amplitudes to higher order, it is known that the $W$
bosons reggeize, and (\ref{2to2born}) takes the form: 
\beq
A^{LLA}=A^{(0)} \,\frac{1}{2}\left( \left( \frac{s}{M^{2}}\right)
^{\omega (t)}+\left( - \frac{s}{M^{2}}\right) ^{\omega (t)}\right)
\,\sim A^{(0)} \left( \frac{s}{M^{2}}\right) ^{\omega(t)},
\label{regge2to2}
\end{equation}
where $\alpha_w (t) = 1+\omega_w(t)$ is the $W$ boson Regge trajectory, and
\beq
\omega_w (t)= (t-M^2) \beta(\bq^2)\,,\,\,\,
\beta(\bq^2) =  g^2 \int \frac{d^{2}k}{(2\pi)^3 }\,
\frac{1}
{\left(\bk^{2}+M^{2} \right)
\left( (\bq-\bk)^{2}+M^{2}\right) }\,,\,\,t=-\bq^{2}.
\eeq

Let us now turn to the unified model of electroweak interactions. Starting
from the nondiagonal mass matrix of the fields $W^{(3)}$ and $B$
\begin{equation}
M^2
\left( W^{(3)}, B \right) \left(
\begin{array}{cc}
1 & -\tan\,\vartheta _{w} \\
-\tan\,\vartheta _{w} & \tan^{2}\,\vartheta _{w}
\end{array}
\right) \left(
\begin{array}{c}
W^{(3)} \\
B
\end{array}
\right) \,,
\label{mass}
\end{equation}
we  introduce their linear combinations
corresponding to  the $Z$
boson and photon:
\beqn
Z&=& c_wW^{(3)}-s_wB \nonumber\\
A&=&s_w W^{(3)}+ c_w B\,, \eeqn where \beq \tan \,\vartheta
_{w}=\frac{g^{\prime }}{g}\;\;, c_w = \cos \,\vartheta _{w}\;\;,
s_w = \sin \vartheta _{w}. \eeq In the new basis, the mass matrix
becomes diagonal with the eigenvalues \beq M_{Z}^2 =
\frac{M^2}{c_w^2}\,,\;\;\;M_{\gamma}^2=0\,. \eeq In the following
we will put $M_W=M$ and $W^{(\pm)} = \frac{1}{\sqrt{2}} \left(
W^{(1)} \mp i W^{(2)}\right)$.

With these definitions we generalize our previous results of the
$SU(2)$ spontaneously broken gauge theory to the Weinberg-Salam
model. Starting from the ($W^{(3)},B$) basis, the propagator of
the neutral bosons can be written in the following operator form:
\begin{equation*}
D_{\mu \nu }(k)=\left(
\begin{array}{c}
c_{w} \\
-s_{w}
\end{array}
\right) \frac{\delta _{\mu \nu }-\frac{k_{\mu }k_{\nu }}{M_{z}^{2}}}{%
k^{2}-M_{z}^{2}}\left(
\begin{array}{cc}
c_{w} \,, & -s_{w}
\end{array}
\right) \end{equation*} \beq +\left(
\begin{array}{c}
s_{w} \\
c_{w}
\end{array}
\right) \frac{\delta _{\mu \nu }}{k^{2}}\left(
\begin{array}{cc}
s_{w} \,, & c_{w}
\end{array}
\right) ,
\eeq
where we have used the physical gauge for the $Z$-boson
and the Feynman gauge for the photon.

The linearized interaction of these vector bosons with the Higgs
field $\varphi$, in the $(W^{(3)},B)$-representation, contains the matrix
\beq
g\,M\,\left(
\begin{array}{cc}
1 & -\tan\,\vartheta _{w} \\
-\tan\,\vartheta _{w} & \tan^{2}\,\vartheta _{w}
\end{array}
\right),
\label{Higgsvert}
\eeq
proportional to the mass matrix (\ref{mass}).
In the $(Z,A)$-representation this matrix becomes diagonal with
only one non-zero coupling constant, $gM_{z}^{2}/M_W$, for the
$ZZ\varphi$-interaction.

As to the other interaction terms, we first note that, when working
in the leading $\ln s$ approximation, and restricting ourselves to scattering
processes of vector bosons and Higgs particles, we can disregard the fermions.
As a result, in the $(W^{(3)},B)$-representation, the
$U(1)$ gauge boson, $B$, decouples and only the $W^{(3)}$-boson enters in the
Yang-Mills action (together with $W^{(\pm )}$-bosons). Therefore, all
gauge boson interaction terms that are needed for our discussion are
obtained by starting from the $SU(2)$ part of the Yang-Mills action and
substituting
\beq
W^{(3)}=c_{w}Z+s_{w}A\,.
\label{mixing}
\eeq

We now turn to the $2 \to 2$ scattering amplitudes, eqs.(\ref{2to2born}),
(\ref{2to2bornhiggs}).
For the vector exchange propagators we replace
\begin{equation}
\frac{1}{\bq^2 +M^2}\rightarrow \frac{c_w^2}
{\bq^2+M_Z^2}\;,\;\;\;\frac{s_w^2}{\bq^2}\,.
\label{propagator}
\end{equation}
for $Z$ and $\gamma$ exchanges, respectively. For the helicity
conserving couplings, $a_{\lambda}$ , we have to observe that the masses
of external and exchanged vector bosons can be different from each other.
Therefore, repeating and generalizing the algebra outlined in
Ref.~\cite{Lip76} one finds for the helicity factor of longitudinally vector bosons:
\begin{figure}
\begin{center}
\epsfig{file=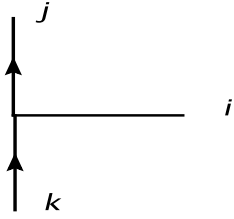,width=3cm,height=3cm}
\caption{Mass assignment in the reggeon - particle - particle vertex}
\end{center}
\end{figure}
\beq
a_3^{i;jk} = \frac{M_i^2-M_j^2-M_k^2}{2M_j M_k}\,,
\label{helicity}
\eeq
where the labels $i,j,k$ refer to exchanged, outgoing vector, and incoming 
particle, respectively (Fig.1). For the transverse polarization the helicity 
factors $a_{\lambda}$ remain the same as in the pure $SU(2)$ case.
As before, each helicity factor $a_{\lambda}$ is multiplied by a helicity 
conserving Kronecker $\delta$-function, e.g. 
$\delta_{\lambda_A,\lambda_{A'}}$.  
Using the labels $W$, $\gamma$, $Z$, $W_3$, we define new helicity factors 
$a_{\lambda}^{i;jk}$, which include, in addition to the pure helicity part 
in eq.(\ref{helicity}), also 
the Kronecker delta functions and the isospin factors, $T^c_{A'A} = 
-i \epsilon_{cA'A}$. In the basis of the charged $W$ bosons, we have 
$T^{W_3}_{W^{(+)} W^{(-)}} = - T^{W_3}_{W^{(-)} W^{(+)}} = +1$ 
(in the lower indices, the first one refers to the final state, the second 
one to the initial state; we count all particles as incoming); 
each permutation or charge conjugation introduces a change in sign. 
Finally, we have to include the coefficients $c_w$, $s_w$ from 
(\ref{mixing}).
We summarize the results for these reggeon-particle-particle 
couplings in Table 1 (we still use the same letter
$a_{\lambda}^{i;jk}$ as in (\ref{helicity})). Here we have listed only
those configurations for which the isospin
factors are $+1$. The other configurations can be obtained by observing the 
antisymmetry of the isospin factors; for example,
\beq
a_{\lambda}^{Z; W^{(-)}W^{(+)}} = - a_{\lambda}^{Z; W^{(+)}W^{(-)}},\;\;
a_{\lambda}^{W^{(-)};Z W^{(+)}} = - a_{\lambda}^{W^{(-)};W^{(+)}Z}\,.
\label{asymmetry}
\eeq 
Note that, for the $Z$-boson and for the photon, the $t$-channel propagators
include additional factors (see (\ref{propagator})).
$$\begin{array}{|c|c|c|c|c|c|}
\hline
&a_{\lambda}^{W^{(-)};ZW^{(+)}} &
a_{\lambda}^ {W^{(-)};\gamma W^{(+)}} &
a_{\lambda}^{Z;W^{(+)}W^{(-)}} &
a_{\lambda}^{\gamma; W^{(+)}W^{(-)}} &
a_{\lambda}^{W_3;W^{(+)}W^{(-)}}\\
\hline
\lambda = 1,2    & -c_w & -s_w & -1 & -1 & -1  \\ \hline
\lambda = 3 &  - \frac{1}{2} &0 &  -1+\frac{1}{2c_w^2} &-1 &-\frac{1}{2}
\\ \hline
\end{array}$$
$$\begin{array}{|c|c|c|c|c|}
\hline
& a_{\lambda}^{W;HW} &
a_{\lambda}^{Z;HZ} &
a_{\lambda}^{\gamma;H \gamma} & a_{\lambda}^{W_3;HZ}\\
\hline
\lambda = 1,2    & 0 &0 &0&0\\ \hline
\lambda = 3 &  -\frac{1}{2}
& -\frac{1}{2c_w^2} & 0 & -\frac{1}{2}  \\ \hline
\end{array}$$
\begin{center}
Table 1: reggeon - particle - particle couplings
\end{center}

As a result, the $2 \to 2$ Born amplitude for the process
$ii' \to kk'$ with the exchange of boson $j$ has the general form:
\beq
A^{(0)} = 2 s g a_{\lambda_i}^{j;i'i} \frac{1}{-\bq^2 -M_j^2}
g a_{\lambda_{i'}}^{j;k'k}\,,
\label{Born}
\eeq
with the substitution (\ref{propagator}) for $Z$ and $\gamma$ exchanges,
and the couplings $a_{\lambda_{A}}^{i;jk}$ have to be read off from the
table. This completes the generalization of
eqs.(\ref{2to2born}) and (\ref{2to2bornhiggs}) to the Weinberg-Salam model.

A final remark on expression (\ref{regge2to2}). In the pure
$SU(2)$ case we know that the $W$ bosons reggeize, which means
that the form (\ref{regge2to2}) is valid. For the Weinberg-Salam
theory, however, we have to find which of the bosons reggeize. It
will turn out that in the neutral channel neither the $Z$-boson
nor the photon lie on Regge trajectories (see also Refs.
~\cite{GS79, LS}). As a result, the simple expression
(\ref{regge2to2}) is valid only for the exchange of charged vector
mesons, but not for the neutral vector exchange.

It will be useful to introduce a convenient diagrammatic notation.
Since neutral and charged vector bosons are behaving quite
differently, it will be helpful to distinguish between them: solid
lines will be used to denote the charged $W$-boson propagators,
and wavy lines stand for the neutral particle propagators (note,
however, that only the $Z$ part of the corresponding matrix
(\ref{Higgsvert}) couples to the Higgs boson). Examples are given
in Fig.2.
\begin{figure}
\begin{center}
\epsfig{file=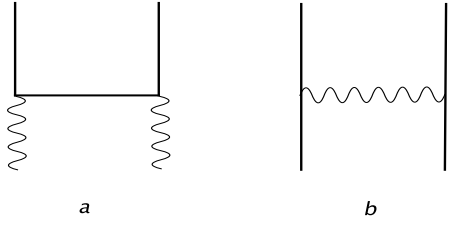,width=6cm,height=3cm} \caption{Two body
scattering processes (black lines denote charged bosons, wavy
lines stand for neutral bosons): (a) $ZZ \to W^{(+)} W^{(-)}$; (b)
$W^{(+)}W^{(-)} \to W^{(+)}W^{(-)}$.}
\end{center}
\end{figure}
\section{Production Amplitudes}
\begin{figure}
\begin{center}
\epsfig{file=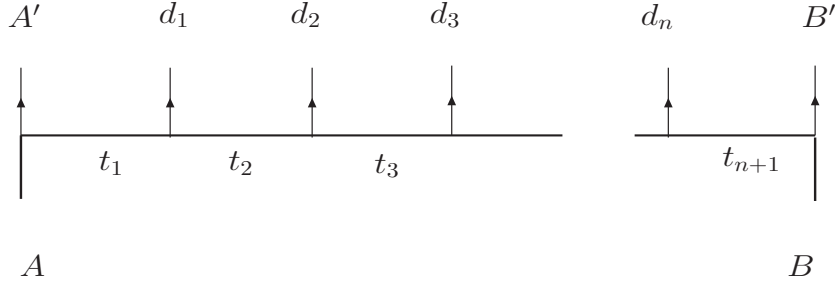,width=12cm,height=4cm}
\caption{Production process $A+B \to A'+d_1+d_2...+d_n+B'$}
\end{center}
\end{figure}
Let us now consider production amplitudes $A+B\rightarrow A^{\prime }+B^{\prime
}+d_{1}+d_{2}+...+d_{n}$ (where $k_{0}=p_{A^{\prime
}},\,k_{n+1}=p_{B^{\prime }}$) in the multiregge region:
\beq
s\gg s_{i}=(k_{i-1}+k_{i})^{2}\gg \bq_r^{2}=\left(
\bp_A -\sum_{i=0}^{r-1} \bk_i\right)
^{2}\sim M^{2}\sim M_{h}^{2}\,.
\label{multiregge}
\eeq
We again begin with the pure $SU(2)$ case. In the Born approximation
the production amplitude equals:
$$
A_{AB}^{(0)\;A^{\prime }B^{\prime }d_{1}...d_{n}}=
$$
\beqn
\label{2tonborn}
2s\,\,g\,a_{\lambda
_{A}}\delta _{\lambda _{A},\lambda _{A^{\prime }}}T_{A^{\prime }A}^{c_{1}}\,%
\frac{1}{-\bq_1^2-M^{2}}\;
\,gT_{c_{2}c_{1}}^{d_{1}}C_{\mu _{1}}(q_{2},q_{1})e^{\mu _{1}}(k_{1})%
\frac{1}{- \bq_2^{2}-M^{2}}...g\,a_{\lambda _{B}}\delta
_{\lambda _{B},\lambda _{B^{\prime }}}T_{B^{\prime }B}^{c_{n+1}}\,,
\eeqn
where the effective vertex $C_{\mu
}(q_{r+1},q_{r})\,,\,\,k_{r}=q_{r}-q_{r+1} $ for $r=1$ is given by:

\begin{equation}
\label{effvertex}
C(q_{2},q_{1})=-q_{1\;\perp}-q_{2\;\perp}-p_{A}\left(
\frac{\bq_{1}^2+M^{2}}{k_{1}p_{A}}-\frac{k_{1}p_{B}}{p_{A}p_{B}}\right)
+ p_{B} \left(
\frac{\bq_2^2 + M^{2}}{k_{1}p_{B}}-\frac{k_{1}p_{A}}{p_{A}p_{B}}\right).
\end{equation}
It has the simple Ward identity property
\begin{equation}
\label{ward}
k_1^{\mu}C_{\mu}(q_2,q_1)=0\,,
\end{equation}
where $k_1=q_1-q_2$, and we have used the reality condition
\beq
\frac{2\,k_{1}p_{A}\,2\,k_{1}p_{B}}{s}=\bk_1^2+M^2
\label{mass-shell}
\eeq
for the produced particle.

In the case where, instead of a $W$-boson with the momentum $k_{r}$, a
Higgs particle with the momentum $k_{r}$ is produced, we substitute
\beq
C_{\mu _{r}}(q_{r+1},q_{r})\,e^{\mu
_{r}}(k_{r})\,T_{c_{r+1}c_{r}}^{d_{r}}\rightarrow M\,\delta
_{c_{r+1}c_{r}}\,.
\label{effvertexhiggs}
\eeq

When in LLA higher order corrections are taken into account,
the production amplitude, in the pure $SU(2)$ case, has the multi-Regge
form (neglecting signature factors):
\beq
A_{2\rightarrow 2+n}^{LLA}=A_{2\rightarrow 2+n}^{(0)}\,\left( \frac{s_{1}}{%
M^{2}}\right) ^{\omega (t_{1})}\left( \frac{s_{2}}{M^{2}}\right) ^{\omega
(t_{2})}...\left( \frac{s_{n+1}}{M^{2}}\right) ^{\omega
(t_{n+1})}\,,\,\,\,s_{r}=2k_{r-1}k_r\,,\,\,t_{r}=-\bq_r^{2}\,.
\label{regge2ton}
\eeq
(as we will see below, for the Weinberg-Salam model the generalization
of the Born amplitude will be slightly more complicated). To apply the
$s$-channel unitarity
one needs to know
the product of two effective vertices $C_{\mu}$. Using the
mass shell condition (\ref{mass-shell})
we obtain:
\beqn
\label{squarevertex}
C_{\mu }(q_{2},q_{1})\,C^{\mu }(q-q_{2},q-q_{1})=\hspace{9cm}
\nonumber \\
2\;\; \frac{(\bq_1^2+M^2)((\bq-\bq_2)^2+M^2)+(\bq_2^2+M^2)
((\bq -\bq_1)^2 +M^2)} {(\bq_1 -\bq_2)^2+M^2} -
2\; \bq^2-3 \; M^{2}\,.
\label{prodC}
\eeqn

One should also calculate the product of two isospin matrices.
We decompose them in terms of various isospin structures in the $t$-channel:
\beq
\varepsilon _{ABd\,}\,\,\varepsilon _{A^{\prime}B^{\prime }d} = 
\sum_{T=0,1,2} r^{(T)} P_{AB}^{A^{\prime }B^{\prime }}(T),
\;\;\;r^{(0,1,2)}=(2,1,-1)\,.
\label{SU2decomp}
\eeq
Here $P_{AB}^{A^{\prime }B^{\prime }}(T)$ are the projectors to the
isospin states with $T=0,1,2$:
\beqn
P_{AB}^{A^{\prime }B^{\prime }}(0)&=&\frac{1}{3}\delta _{AA^{\prime }\,}\delta
_{BB^{\prime }\,},\nonumber \\
P_{AB}^{A^{\prime }B^{\prime }}(1)&=&\frac{1}{2}%
\varepsilon _{cA^{\prime }A\,}\varepsilon _{cB^{\prime }B\,},\nonumber \\
P_{AB}^{A^{\prime }B^{\prime }}(2)&=&\frac{1}{2}\left( \delta
_{AB}\,\delta _{A^{\prime }B^{\prime }\,}+\delta _{AB^{\prime }\,}\delta
_{A^{\prime }B\,}\right) -\frac{1}{3}\delta _{AA^{\prime }\,}\delta
_{BB^{\prime }\,}.
\eeqn

Let us now turn to the realistic Weinberg-Salam model. The main task is the
generalization of the effective production vertex (\ref{effvertex})
to the case where the attached $t$-channel bosons have different masses
($M_W$ for the $W$ boson, $M_Z$ for the $Z$ boson, or zero mass for the
photon). Again, it is needed to return to Ref. ~\cite{Lip76} for computing
the $2 \to 3$ production amplitudes in the double Regge limit.
The result of this analysis which will not be presented in detail is that the 
Born approximation is still of the factorized form
(\ref{2tonborn}), where the couplings to the incoming particles, 
$a_{\lambda}^{i;jk}$, are the same as in Table 1. 
In the crossing channels we have the propagators
$1/(- \bq_i ^2 - M_i^2)$. If we denote the masses of the exchanged vector
particle on the right (left) hand side of a produced vector boson
with mass $M$ by $M_2$ ($M_1$), the effective production vertex becomes
\beq
\label{effvertex1}
C(q_{2},q_{1})^{M; M_2 M_1}=-q_{1\;\perp}-q_{2\;\perp}-p_{A}\left(
\frac{\bq_{1}^2+M_1^{2}}{k_{1}p_{A}}-\frac{k_{1}p_{B}}{p_{A}p_{B}}\right)
+ p_{B} \left(
\frac{\bq_2^2 + M_2^{2}}{k_{1}p_{B}}-\frac{k_{1}p_{A}}{p_{A}p_{B}}\right)
\eeq
(note that the dependence upon $M$ is through eq.(\ref{mass-shell})).
If the produced vector particle is a $Z$ boson (photon), an additional
factor $c_w$ ($s_w$) has to be included. Finally, each exchanged $Z$ boson
receives, in the numerator, a factor $c_w^2$, each photon propagator a factor
$s_w^2$ (see (\ref{propagator})). For the Higgs production we can use
 (\ref{effvertexhiggs}),
where on the rhs $M$ becomes $M_W$, if the Higgs is produced from
$W^{(\pm)}$ exchange. For Higgs production from a $Z$ exchange, replace
$M \to M_W/c_w^4$ (and retain the factor $c_w^2$ for each $Z$ exchange
propagator). Finally, in (\ref{regge2ton}) the group factors 
$T^{d_1}_{c_2c_1}$ have to be rewritten in terms of $Z$ and $W^{(\pm)}$ 
(cf. the discussion before (\ref{asymmetry}); note that both the $Z$ and the 
photon couple to the third component of the isospin generator: 
$T^{Z}_{c_2c_1} = T^{\gamma}_{c_2c_1}=T^{3}_{c_2c_1}$).  

The Ward identity (\ref{ward}) for the production vertex is replaced now by
the relation
\begin{equation}
\label{ward1}
k_1^{\mu}C_{\mu}^{M; M_2 M_1}(q_2,q_1)= M_2^2 - M_1^2\,.
\end{equation}
For the $s$-channel unitarity integration we again need the product of two
effective vertices. More precisely, one should sum over the physical helicities
of the produced boson with mass $M_m$:
\beq
  \label{eq:sum2}
  \sum_\lambda \epsilon_\lambda^\mu (k_1) \epsilon_\lambda^\nu
  (k_1)=-g^{\mu\nu}+\frac{k_1^\mu k_1^\nu}{M_m^2}
\eeq
(note that for the production of a photon with $M_m=0$ the second term is
absent in an accordance with the vanishing of (\ref{ward1}) for $M_1=M_2$).
\begin{figure}
\begin{center}
\epsfig{file=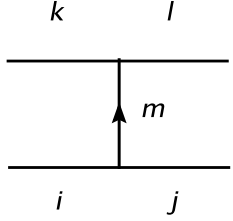,width=3cm,height=3cm}
\caption{Assigment of masses for the product of two effective vertices}
\end{center}
\end{figure}
For the mass assignment illustrated in Fig.4 we obtain 
(cf.(\ref{squarevertex})):
\beqn
\label{squarevertex1}
&&\hspace{-1cm}C_{\mu}^{M_m; M_j M_i}(q_j,q_i)\,(g^{\mu\nu}-
\frac{k_1^\mu k_1^\nu}{M_m^2})\,
C_{\nu}^{M_m; M_l M_k}(q-q_j,q-q_i)=
  \\ \nonumber \\
\nonumber
&&-2 \bq^2 +M_m^2-M_i^2-M_j^2-M_k^2-M_l^2 +
\frac{(M_j^2-M_i^2)(M_l^2-M_k^2)}{M_m^2}+\\
&&2\; \frac{(\bq_i^2+M_i^2)((\bq-\bq_j)^2+M_l^2)
+ ((\bq-\bq_i)^2 + M_j^2)(\bq_j^2 + M_k^2)}{(\bq_i-\bq_j)^2+M_m^2}
  \nonumber \, .
\eeqn
This result can be obtained with the use of eqs.(\ref{mass-shell}) 
(\ref{effvertex1}), and (\ref{ward1}).

\section{One and two loop results}
\subsection*{$2 \to 2$ scattering in one loop}
We are now ready to carry out the BFKL program. Beginning with
one loop amplitudes, we first consider the charged isospin-1
exchange. To be definite, let us study the process $ZZ \to
W^{(+)}W^{(-)}$.
\begin{figure}
\begin{center}
\epsfig{file=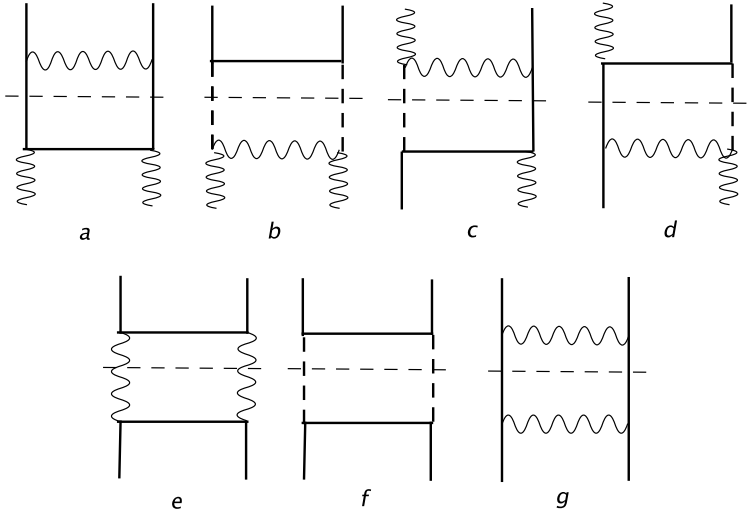,width=10cm}
\caption{One loop corrections to $2 \to 2$ -- processes shown in Fig.2}
\end{center}
\end{figure}
The Born diagram is shown in Fig.2a, the first corrections come from the
box diagrams of the type Fig.5a-d. For the energy discontinuities
we use the unitarity conditions, e.g.
\beqn
  \label{1loopvirt2}
  {\rm Im_s}\, A^{(1)}_{ab \to a^\prime b^\prime}=\frac 12 \sum_i \int d \Pi_2
 A^{(0)}_{ab\to i}(k^2) A^{(0)\dagger}_{i\to a^\prime
 b^\prime}((k-q)^2) \,,
\eeqn
where the sum in $i$ extends over all possible intermediate
two-particle states, and we then make use of dispersion relations
to compute the scattering amplitudes. We define
the functions $\beta_{ij}(q^2)$:
\begin{equation}
  \label{beta}
  \beta_{ij}(q^2)=
 g^2 \int \frac{d^2 k}{(2\pi)^3} \frac1{\bk^2+M_i^2} \,
  \frac1{(\bk-\bq)^2+ M_j^2}\, .
\end{equation}
We also use their generalizations:
\begin{equation}
  \label{beta3}
  \beta_{ijk}(q^2)=
 g^4 \int \frac{d^2k d^2k'}{(2\pi)^6}  \frac1{\bk^2+M_i^2} \,
          \frac1{{\bk'}^2+M_j^2} \,
  \frac1{(\bk +\bk'-\bq)^2+ M_k^2}\, .
\end{equation}
The subscripts indicate the type of vector particles inside the
$\beta$ functions.

The Born amplitude has the form
\beq A^{(0)}_{11} = 2 s g
a_{\lambda_A}^{W^{(+)};W^{(-)}Z} \frac{1}{-\bq^2 - M_W^2} g 
a_{\lambda_B}^{W^{(-)};W^{(+)}Z}.
\label{chargeBorn}
\eeq 
Next we form signatured amplitudes. In our case, $ZZ \to
W^{(+)}W^{(-)}$, they are defined by the combinations 
\beq
\frac{1}{2}\left( A_{ZZ \to W^{(+)}W^{(-)}} \;\;\pm \;\;
A_{W^{(-)}Z\to ZW^{(-)}} \right). 
\eeq 
Signature describes the symmetry under $s \to -s$. Because of the antisymmetry
properties of the isospin coefficients, the Born amplitude for our
process belongs to odd signature (in terms of isospin, it is the
antisymmetric $T=1$ representation, $T_3= \pm 1$). Using the unitarity 
relations for the processes $ZZ \to W^{(+)}W^{(-)}$ (Fig.5a, b) and for the cross 
process $W^{(-)}Z\to ZW^{(-)}$ (Fig.5c, d), we obtain for the odd-signature 
amplitude:   
\beq
  \label{eq:111}
A^{(1)}_{11}=[s\ln (-s)-u\ln(-u)]\, g a_{\lambda_A}^{W^{(+)};W^{(-)}Z}
  \Big\{ c_w^2\,
\beta_{wz}(q^2)+s_w^2\, \beta_{w\gamma}(q^2)\Big\} g  
a_{\lambda_B}^{W^{(-)};W^{(+)}Z} \,.
\eeq
In the LLA approximation the energy scale in the logarithm
is arbitrary; it is natural to chose the scale to be of the order of $M_W$.
We omit to explicitly write this scale.

Comparing the Born approximation with the one loop result,
one is lead to interpret both expressions as being the first two terms
in the power series expansion of (cf.(\ref{regge2to2})) 
\beq 
A_{11} = - g
a_{\lambda_A}^{W^{(+)};W^{(-)}Z} \frac{(-s)^{\alpha_c(q^2)} 
- (-u)^{\alpha_c(q^2)}}
{-\bq^2 - M_W^2} g 
a_{\lambda_B}^{W^{(-)};W^{(+)}Z}
\label{chargeallorder}
\eeq
with the trajectory
function
\beq
\label{chargetrajectory}
\alpha_c(q^2)=1+(q^2-M_W^2)\Big[c_w^2\, \beta_{wz}(q^2)+s_w^2\,
\beta_{w\gamma}(q^2)\Big].
\eeq
This is consistent with the expectation that the charged $W$ bosons reggeize,
in the same way as the $W$ bosons do in the pure $SU(2)$ theory. The same
conclusion holds, if we replace external vector bosons by Higgs bosons.
Later on we will confirm that the reggeization of the charged $W$ bosons
is correct to all orders. It will be convenient to introduce 
\beq
\omega_c(q^2) = \alpha_c(q^2) - 1 =
(q^2-M_W^2)\Big[c_w^2\, \beta_{wz}(q^2)+s_w^2\,
\beta_{w\gamma}(q^2)\Big].
\label{trajWpm}
\eeq

Turning next to the neutral exchange, we consider the elastic scattering of
two charged bosons, the process $W^{(+)}W^{(-)} \to W^{(+)}W^{(-)}$.
The Born diagram is shown in Fig.2b; the amplitude has the
form:
\beq
\label{neutralborn}
A^{(0)}_{10} = 2 s \left( g a_{\lambda_A}^{Z;W^{(-)}W^{(+)}} \frac{c_w^2}
{-\bq^2 - M_Z^2} g a_{\lambda_B}^{Z;W^{(+)}W^{(-)}} 
+ g a_{\lambda_A}^{\gamma;W^{(-)}W^{(+)}}
\frac{ s_w^2} {-\bq^2} g a_{\lambda_B}^{\gamma;W^{(+)}W^{(-)}} \right).
\eeq
It belongs to odd-signature (the $T=1$ representation), and it represents the
neutral, $T_3=0$, component. For the one-loop odd-signature contribution 
we obtain (Figs.5e - f): 
\begin{equation}
  \label{1loopneutral}
A^{(1)}_{10}=[s\ln (-s)-u\ln(-u)]\, g a_{\lambda_A}^{W_3;W^{(-)}W^{(+)}} \;
\beta_{ww}(q^2)  \: g a_{\lambda_B}^{W_3;W^{(+)}W^{(-)}}.
\end{equation}
An analogous result is obtained for the process $ZW^{(-)} \to HW^{(-)}$,
with the substitution $a_{\lambda_A}^{W_3;W^{(-)}W^{(+)}}
\to a_{\lambda_A}^{W_3;HZ}$.  
At this stage, it seems premature to draw any conclusion about the connection
of the one loop result with the Born approximation.

The one-loop even
signature contribution of Fig.5g contributes to both isospin 0 and 2.
We present the sum of both:
\begin{align}
\label{vacuum1loop}
A^{(1)}_{even} = 2 i \pi s \left( -\frac{1}{2} g a_{\lambda_A}^{W_3;W^{(-)}W^{(+)}} \;\beta_{ww}(q^2)\;
g a_{\lambda_B}^{W_3;W^{(+)}W^{(-)}}  \hspace{6cm} 
\right. \nonumber \\ \left.
+ \;g (a_{\lambda_A}^{Z;W^{(-)}W^{(+)}})^2\;  c_w^4\; \beta_{zz}(q^2)\; 
g (a_{\lambda_B}^{Z;W^{(+)}W^{(-)}})^2 
 +\; g  (a_{\lambda_A}^{\gamma;W^{(-)}W^{(+)}})^2 \; s_w^4\; \beta_{\gamma \gamma}(q^2) \;
 g (a_{\lambda_B}^{\gamma;W^{(+)}W^{(-)}})^2 
\right. \nonumber\\ \left.
+\;2 \;g a_{\lambda_A}^{Z;W^{(-)}W^{(+)}} a_{\lambda_A}^{\gamma;W^{(-)}W^{(+)}} \; c_w^2 s_w^2 \;
\beta_{\gamma z}(q^2)\; g
a_{\lambda_B}^{Z;W^{(+)}W^{(-)}} a_{\lambda_B}^{\gamma;W^{(+)}W^{(-)}}  
\right)\,.
\hspace{1.5cm}
\end{align}

\subsection*{Two loop results for $2 \to 2$ scattering}
Two loop corrections consist of two classes of terms, the
two-particle intermediate states and the three-particle
intermediate states in the $s$-channel ~\cite{Lip76}. The former
ones are obtained by inserting, into the bilinear unitarity
relation, the Born term on one side and one loop amplitudes on the
other side. For the calculation of the three particle state we
make use of expression (\ref{squarevertex1}); we also include the
production of Higgs scalars. Let us  begin with the charge
exchange channel. Making use of the vertices in Table 1
and of the one-loop results listed above, and summing over all
2-particle intermediate states, we obtain for the process $ZZ \to
W^{(+)}W^{(-)}$: 
\beq - 2 s \ln^2 s \,g a_{\lambda_A}^{W^{(+)};W^{(-)}Z} \left(
\beta_{www}(\bq^2)
     + c_w^4 \beta_{wzz}(\bq^2)
     +  2 c_w^2 s_w^2 \beta_{wz\gamma}(\bq^2)
     + s_w^4  \beta_{w\gamma\gamma}(\bq^2) \right) g 
a_{\lambda_B}^{W^{(-)};W^{(+)}Z}\,.
\label{charged2part}
\eeq
For the sum over 3-particle intermediate states we find, making use of 
eq.(\ref{squarevertex1}), a sum of two terms. The first
one is:
\begin{align}
(s \frac{\ln^2 (-s)}{2!} - u \frac{\ln^2 (-u)}{2!}) 
           g a_{\lambda_A}^{W^{(+)};W^{(-)}Z}
\left(s_w^2 \beta_{w\gamma}(\bq^2) +  c_w^2 \beta_{wz}(\bq^2) \right)
\nonumber \\
\left(-\bq^2 - M_W^2\right)
\left(s_w^2 \beta_{w\gamma}(\bq^2) +  c_w^2 \beta_{wz}(\bq^2) \right)
g a_{\lambda_B}^{W^{(-)};W^{(+)}Z}.
\end{align}
The second one can be written in the form:
\beq
2 s \ln^2 s \,g a_{\lambda_A}^{W^{(+)};W^{(-)}Z} \left( \beta_{www}(\bq^2)
     + c_w^4 \beta_{www}(\bq^2)
     +  2 c_w^2 s_w^2 \beta_{wz\gamma}(\bq^2)
     + s_w^4  \beta_{w\gamma\gamma}(\bq^2) \right) g 
a_{\lambda_B}^{W^{(-)};W^{(+)}Z}\,.
\eeq
and cancels the entire 2-particle intermediate state, eq.(\ref{charged2part}).
Hence the two-loop result for the negative signature charge exchange channel 
coincides with the second term in the expansion of
\beq
 - g a_{\lambda_A}^{W^{(+)};W^{(-)}Z}  \frac{(-s)^{\alpha_{c}(q^2)}
- (-u)^{\alpha_{c}(q^2)}}
{-\bq^2 -M_W^2}  g a_{\lambda_B}^{W^{(-)};W^{(+)}Z},
\eeq
confirming the reggeization in the one-loop approximation.

Turning to the neutral exchange channel, we again first consider the 
two-particle
intermediate states. For the process $  W^{(+)} W^{(-)}  \to W^{(+)} W^{(-)}$ we obtain,
after summation over all possible 2-particle intermediate states:
\beq
- 2 s \ln ^2 s \,g a_{\lambda_A}^{W_3;W^{(-)}W^{(+)}} \left(s_w^2 \beta_{\gamma ww}
+ c_w^2 \beta_{zww} \right) g a_{\lambda_B}^{W_3; W^{(+)}W^{(-)}},
\label{neutral2part}
\eeq
where the couplings $a_{\lambda}^{W_3;W^{(-)}W^{(+)}}$ are listed in Table 1.
The calculation of the three particle intermediate state, again, makes use of the
square of the production
vertex, eq.(\ref{squarevertex1}). Summing over all possible 3-particle
intermediate states we obtain
a sum of two terms. The first one is:
\beq
(s \frac{\ln^2 (-s)}{2!} - u \frac{\ln^2 (-u)}{2!})\, 
g \,a_{\lambda_A}^{W_3;W^{(-)}W^{(+)}}
\,\beta_{ww}(\bq^2)\, (-\bq^2 - M_W^2)\,  \beta_{ww}(\bq^2)\,
g \,a_{\lambda_B}^{W_3;W^{(+)}W^{(-)}},
\label{neutral2loop}
\eeq
the second one
\beq
2 s \ln^2s\, g\, a_{\lambda_A}^{W_3;W^{(-)}W^{(+)}}
\left(s_w^2 \beta_{\gamma ww}(\bq^2) + c_w^2 \beta_{zww}(\bq^2) \right)
\,g \,a_{\lambda_B}^{W_3;W^{(+)}W^{(-)}}.
\eeq
This second terms cancels against the two-particle contribution,
eq.(\ref{neutral2part}). We have thus only the first term,
(\ref{neutral2loop}), which can be
interpreted as the second term in the expansion of the expression
\beq
 - g a_{\lambda_A}^{W_3;W^{(-)}W^{(+)}}  \frac{(-s)^{\alpha_{n}(q^2)}
- (-u)^{\alpha_{n}(q^2)}}
{-\bq^2 -M_W^2}  g a_{\lambda_B}^{W_3;W^{(+)}W^{(-)}}
\label{reggeneutral}
\eeq
with
\beq
\alpha_{n} (q^2)=1+(q^2 - M_W^2) \beta_{ww}(q^2)\,.
\label{trajW3}
\eeq
In the following we will also use the notation
\beq
\omega_{n}(q^2) = \alpha_{n} (q^2) - 1 = (q^2 - M_W^2) \beta_{ww}(q^2)\,.
\eeq
The expression (\ref{reggeneutral}) matches the one-loop result, 
(\ref{1loopneutral}), but it does not agree with the Born approximation,
(\ref{neutralborn}). We therefore make the following ansatz for the neutral exchange in the $2 \to 2$ scattering process:
\begin{align}
A_{10} = 2 s \left( g a_{\lambda_A}^{Z;W^{(-)}W^{(+)}} \frac{c_w^2}
{-\bq^2 - M_Z^2} g a_{\lambda_B}^{Z;W^{(+)}W^{(-)}} + 
g a_{\lambda_A}^{\gamma;W^{(-)}W^{(+)}}
\frac{ s_w^2} {- \bq^2} g a_{\lambda_B}^{\gamma;W^{(+)}W^{(-)}} \right)
\nonumber \\
- g a_{\lambda_A}^{W_3;W^{(-)}W^{(+)}}  \left( \frac{(-s)^{\alpha_{n}(q^2)} 
-(-u)^{\alpha_{n}(q^2)}}
{-\bq^2 -M_W^2} + \frac{2s}{-\bq^2 -M_W^2}\right) 
g a_{\lambda_B}^{W_3;W^{(+)}W^{(-)}},
\label{neutralallorder}
\end{align}
i.e. we have a Regge pole in the neutral exchange channel,
which passes through unity at $t=M_W^2$: neither the $Z$ boson nor the photon
lie on this trajectory. Note that, in the second line of 
(\ref{neutralallorder}), the pole at $q^2 = M_W^2$ cancels.
For $s_w=0$, we have $M_Z=M_W$, and 
$\alpha_n$ passes through the $Z$-boson.
Later on we shall verify that this result is
correct to all orders.

\subsection*{One loop results for $2 \to 3$ production amplitudes}

Before we can start to write integral equations we need to calculate
corrections to the production
amplitudes: this will be done in the spirit of ~\cite{Ba79}.
To be definite, the process $W^{(+)} W^{(-)} \to ZZZ$ (Fig.6a)
\begin{figure}
\begin{center}
\epsfig{file=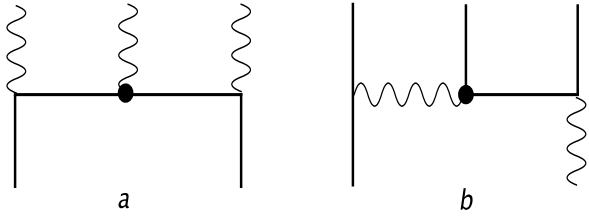,width=8cm,height=3cm}
\end{center}
\caption{$2 \to 3$ production amplitudes (the dot marks the effective
production vertex (eq.(\ref{effvertex1})): (a) $W^{(+)}W^{(-)} \to ZZZ$; (b)
$W^{(+)}Z \to W^{(+)} W^{(+)}W^{(-)}$.}
\end{figure}
will be considered.
In the Born approximation we have:
\beq
A_{2 \to 3}^{(0)} = 2 s g a_{\lambda_A}^{W^{(-)};ZW^{(+)}} 
\frac{1}{-\bq_1^2-M_W^2}
          g T^{Z}_{W^{(-)}W^{(+)}} C (\bq_2,\bq_1)^{M_Z; M_W M_W}
\frac{1}{-\bq_2^2-M_W^2} g a_{\lambda_B}^{W^{(+)};ZW^{(-)}}\,,
\eeq
where the energy variables $s_i$ have been defined in (\ref{multiregge}).  
In order to be able to compute the discontinuity in $s_2$,
we make a more general ansatz which exhibits the analytic structure
in all three energy variables:
\beq
A_{2 \to 3} = F_L s^{\alpha_1} s_{2}^{\alpha_2 - \alpha_1}
\frac{\xi_{\alpha_1}}{t_1 - M_1^2} \frac{\xi_{\alpha_2 \alpha_1}}
{\alpha_2 - \alpha_1}
+ F_R s^{\alpha_2} s_{1}^{\alpha_1 - \alpha_2}
\frac{\xi_{\alpha_2}}{t_2 - M_2^2} \frac{\xi_{\alpha_1 \alpha_2}}
{\alpha_1 - \alpha_2}\,,
\label{F2analytic}
\eeq
where the signature factors are:
\begin{align}
\xi_{\alpha_1} = e^{-i \pi \alpha_1} - 1,\nonumber \\
\xi_{\alpha_1 \alpha_2} = e^{-i \pi (\alpha_1 - \alpha_2)} + 1\,,
\end{align}
and $\alpha_i= \alpha(t_i)$ are the corresponding Regge trajectories. 
The two partial waves $F_L$ and $F_R$ will be determined from the
discontinuities in the $s_2$ and $s_1$ channels, resp., and we shall see that
the ansatz (\ref{F2analytic}) is compatible with the familiar factorized
multiregge form. The discontinuity in the $s_2$ channel is:
\beq
disc_{s_2} A_{2 \to 3} = - \pi \frac{s^{\alpha_1} s_2^{\alpha_2 - \alpha_1} \xi_{\alpha_1}}{t_1 - M_1^2}
F_L\,.
\eeq
In the lowest order, we simply put $\alpha_i\to 1$ and $\xi_{\alpha_1} \to -2$.

When computing the unitarity integral in the $s_2$-channel (Fig.7)
\begin{figure}
\begin{center}
\epsfig{file=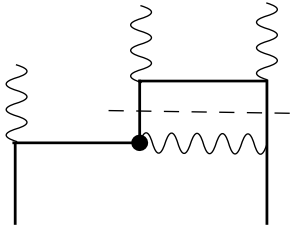,width=4cm,height=3cm}
\end{center}
\caption{Unitarity integral in the $s_2$-subchannel.}
\end{figure}
it is convenient to first transform into the center-of-mass system
of the $s_2$-channel, to multiply with the $2\to2$ scattering
amplitude having a simple helicity structure in the $s_2$ channel,
to compute the two-body phase space integral, and finally to
transform back into the overall cm-system. Details of this
procedure have been described in ~\cite{Ba79}; some of the
formulae, however, have to be generalized to the case of unequal
masses of the vector bosons. A list of the relevant expressions is
presented in the appendix. After summing over all possible
s-channel intermediate states and over all $t$-channel exchanges
we find for the partial wave $F_L$:
\beqn 
F_L= -\frac{1}{2} s g a_{\lambda_A}^{W^{(-)};W^{(+)}Z} \left(
          g T^{Z}_{W^{(-)}W^{(+)}} C (\bq_2,\bq_1)^{M_Z;M_W M_W} 
\frac{\omega_{c}(q_2^2)}
              {-\bq_2^2 -M_W^2} \right.\nonumber \\
\left.
- (\bq_1^2 + M_W^2) (c_w^2K^{Z;WW} +s_w^2 
K^{\gamma;WW}) \right)
   g a_{\lambda_B}^{W^{(+)};W^{(-)}Z}\,.
\eeqn
Here we have introduced the short-hand notation:
\beq
K^{Z;WW} = g^2 T^{Z}_{W^{(-)}W^{(+)}} \int \frac{d^2 k}{(2\pi)^3}
\frac{g C^{M_Z;M_W M_W}(\bq_2-\bk,\bq_1-\bk)}
                     { ((\bq_1-\bk)^2 + M_W^2) ((\bq_2-\bk)^2 + M_W^2)}
                    \frac{1}{(\bk^2 +M_Z^2)}\,.
\eeq
With an analogous result for the discontinuity in $s_1$ and for $F_R$ we
return to (\ref{F2analytic}). In the sum of both partial waves, the
terms containing $K^{Z;WW}$ and $K^{\gamma;WW}$ cancel, and
we are left with the expression
\beq
 A_{2 \to 3} =
2 s a_{\lambda_A}^{W^{(-)};W^{(+)}Z}
          \frac{s_1^{\omega_c(q_1^2)}}{-\bq_1^2-M_W^2}
          g T^{Z}_{W^{(-)}W^{(+)}}C (\bq_2,\bq_1)^{M_Z;M_W M_W}
\frac{s_2^{\omega_{c}(q_2^2)}}{-\bq_2^2-M_W^2} g 
a_{\lambda_B}^{W^{(+)};W^{(-)}Z},
\label{Zproduction}
\eeq
i.e. the $W$ exchanges have started to reggeize. It is straightforward to 
verify, in lowest order, the discontinuities in $s_1$ and $s_2$ which led 
to the partial waves $F_L$ and $F_R$.

In an analogous way we compute the one loop corrections to other production
amplitudes. For the process $W^{(+)}Z \to W^{(+)} W^{(-)}W^{(+)}$ (see Fig.6b) 
\begin{figure}
\begin{center}
\epsfig{file=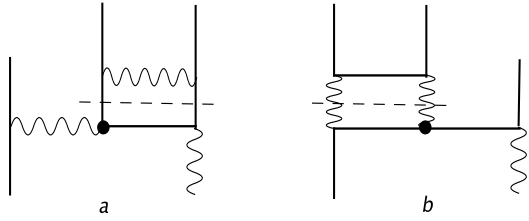,width=7cm,height=3cm}
\end{center}
\caption{Unitarity integrals (a) in the $s_2$ subchannel, (b) in the
$s_1$ channel.}
\end{figure}
we have in the Born approximation:
\begin{align}
A_{2 \to 3} = 2 s 
\left( g a_{\lambda_A}^{Z;W^{(-)}W^{(+)}} \frac{c_w^2}{-\bq_1^2-M_Z^2}
          g T^{W^{(+)}}_{W^{(-)}Z}C (\bq_2,\bq_1)^{M_W; M_W; M_Z}
\right. \nonumber\\ \left.               
+ g a_{\lambda_A}^{\gamma;W^{(-)}W^{(+)}} \frac{s_w^2}{-\bq_1^2}
          g T^{W^{(+)}}_{W^{(-)}\gamma} C (\bq_2,\bq_1)^{M_W;M_W M_{\gamma}} 
\right)
  \cdot\;\; \frac{1}{-\bq_2^2-M_W^2} 
      g a_{\lambda_B}^{W^{(+)};W^{(-)}Z}.
\end{align}

For the $t_2$ channel we expect that, in higher orders, the $W$-exchange 
will reggeize. As to the $t_1$ channel, our analysis of the $2 \to 2$ 
scattering process with neutral exchange, eq.(\ref{neutralallorder}), 
suggests that, in higher order, in addition to the elementary $z$ and 
$\gamma$ exchanges the neutral Regge pole, $\alpha_n$, should appear. 
As we have seen before, Regge pole exchanges contribute to the discontinuities 
in $s_1$ and $s_2$, whereas the elementary 
$z$ and $\gamma$ exchanges do not. Therefore, our ansatz  
(\ref{F2analytic}) with $\alpha_1 \to \alpha_n$, $\alpha_2 \to \alpha_c$ 
should be valid for the Regge pole exchanges in both
crossing channels, but we have to add extra terms for the elementary exchanges 
in the $t_1$ channel which have a discontinuity in $s_2$ but not in $s_1$, 
e.g. for $Z$ exchange:
\beq
F_L^Z s s_2^{\alpha_c -1} \frac{-2}{t_1-M_Z^2} 
\frac{\xi_{\alpha_c 1}}{\alpha_c - 1}\,.
\eeq 

Proceeding in the same way as before we compute, from the single 
discontinuities in $s_1$ and $s_2$, the partial waves $F_L$ and $F_R$ 
(see Fig.8). 
From this we infer the following all-order expression:
\begin{align}
A_{2 \to 3} = 2 s \left( g a_{\lambda_A}^{W_3;W^{(-)}W^{(+)}}
\frac{s_1^{\omega_{n}(q_1^2)}-1}{-\bq_1^2-M_W^2}
          g T^{W^{(+)}}_{W^{(-)}W_3}C (\bq_2,\bq_1)^{M_W;M_W M_W}
  + \right.  \nonumber \\
\left.  g a_{\lambda_A}^{Z;W^{(-)}W^{(+)}} \frac{c_w^2}{-\bq_1^2-M_Z^2}
          g T^{W^{(+)}}_{W^{(-)}Z} C (\bq_2,\bq_1)^{M_W;M_WM_Z}
              + \right.  \nonumber \\
\left.g a_{\lambda_A}^{\gamma;W^{(-)}W^{(+)}} \frac{s_w^2}{-\bq_1^2}
          g T^{W^{(+)}}_{W^{(-)}\gamma}C (\bq_2,\bq_1)^{M_W;M_W M_{\gamma}} 
\right) 
 \cdot\;\;  \frac{s_2^{\omega_{c}(q_2^2)}}{-\bq_2^2-M_W^2}
      g a_{\lambda_B}^{W^{(+)};W^{(-)}Z}.
\label{Wproduction}
\end{align}
In the charge exchange channel ($t_2$-channel) we recognize the reggeization
of the $W$ boson, whereas the neutral exchange channel  ($t_1$-channel)
has the same structure as (\ref{neutralallorder}). In (\ref{Wproduction})
the new element is the $W$ production vertex where one of the reggeons belongs
to the neutral Regge pole, $\alpha_{n}$: its particle pole lies at
$M_W$, and consequently the mass labels of the production vertex are
$C (\bq_2,\bq_1)^{M_W; M_W M_W}$.

As a final example, we calculate the production process
$W^{(+)} W^{(-)} \to W^{(+)} H W^{(-)}$.
Our one-loop calculation leads to:
\beqn
A_{2 \to 3}= 2 s \left(a_{\lambda_A}^{Z;W^{(-)}W^{(+)}} 
\frac{c_w^2}{-\bq_1^2 -M_Z^2} \frac{1}{c_w^2}
+ a_{\lambda_A}^{W_3;W^{(-)}W^{(+)}} \frac{s_1^{\omega_{n}(q_1^2)} - 1}
{-\bq_1^2 -M_W^2} \right) M_W \nonumber \\
\cdot\;\;\left( \frac{1}{c_w^2} \frac{c_w^2}{-\bq_2^2 -M_Z^2} 
a_{\lambda_B}^{Z;W^{(+)}W^{(-)}} +
\frac{s_2^{\omega_{n}(q_2^2)} - 1} {-\bq_2^2
-M_W^2}a_{\lambda_B}^{W_3;W^{(+)}W^{(-)}}\right). 
\label{Higgsproduction}
\eeqn 
Note that, following our convention defined before, 
each $Z$ exchange carries a factor $c_w^2$. As a consequence, the production
of the Higgs obtains a factor $1/c_w^2$ if, 
in (\ref{Higgsproduction}), one of the 
attached neutral exchanges is a $Z$ boson, and a factor $1/c_w^4$ if 
we have a $Z$ boson on both sides of the produced Higgs. 
In the following we shall verify that these production amplitudes 
lead to the correct bootstrap equations.
\section{Integral Equations}

We now turn to the derivation of integral equations which represent the
sum of discontinuities of the scattering amplitude
over an arbitrary number of produced particles.
\begin{figure}
\begin{center}
\epsfig{file=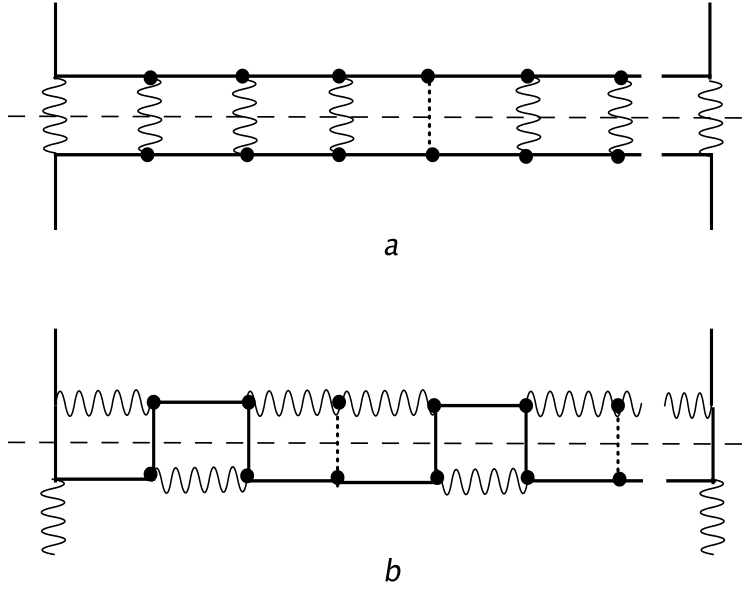,width=10cm,height=8cm}
\end{center}
\caption{Ladder diagrams obtained from the square of production amplitudes:
(a) $ZZ \to W^{(+)}W^{(-)}$ (charge exchange);
(b) $W^{(+)}W^{(-)} \to W^{(+)}W^{(-)}$ (neutral exchange, odd signature).}
\end{figure}
The one and two loop calculations suggest that the $2 \to n$ production
amplitudes can be written in the factorized multiregge form, where
charged and neutral exchanges lead to slightly different expressions. 
The exchange of a charged gauge boson requires the usual reggeon propagator 
$s_i^{1+\omega_c}/(\bq^2+M_W^2)$. 
For the neutral exchange we have a sum of three terms, the $Z$ and $\gamma$
exchange in the Born approximation, and the neutral Regge pole exchange.
In the angular momentum representation, the corresponding propagators are
\beqn
\frac{c_w^2}{\omega}   \frac{1}{\bq^2+M_Z^2},\hspace{1cm}
\frac{s_w^2}{\omega}   \frac{1}{\bq^2},\hspace{1cm}
\left(\frac{1}{\omega- \omega_c(q^2)} - \frac{1}{\omega}\right)
\frac{1}{\bq^2+M_W^2}\,,
\eeqn
resp. When inserting the sum of these three terms into a production
amplitude, each term comes with its own coupling to external and
produced particles.
For example, the couplings to an external $W$ boson are
$a_{\lambda}^{Z;W^{(-)}W^{(+)}}$, $a_{\lambda}^{\gamma;W^{(-)}W^{(+)}}$, 
and $a_{\lambda}^{W_3;W^{(-)}W^{(+)}}$,
respectively. When, inside a multiregge production amplitude, the exchanged 
neutral boson couples to a $W$ production vertex, the effective production
vertices are $C^{M_W;M_Z M_W}$, $C^{M_W;M_{\gamma} M_W}$, and
$C^{M_W; M_3 M_W}$ with $M_3=M_W$, resp. 
With these rules it will be straightforward to write
down the integral equations for the sum of products of production amplitudes 
(cf.~\cite{BFKL2})).

Let us begin with the partial wave representations. For the 
$2 \to 2$ process with charged boson exchange we again consider the 
process $ZZ \to W^{(+)}W^{(-)}$  (eqs.(\ref{chargeBorn}) 
and (\ref{chargeallorder})).
The $t$-channel partial wave decomposition contains the Born contribution 
(\ref{chargeBorn}) and, from the Regge pole,  
the integral over $\omega =j-1$:
\begin{equation}
A_{11}(s,t)= \frac{2s}{-\bq^2 - M_W^2} g a_{\lambda_A}^{W^{(+)};W{(-)}Z} 
g a_{\lambda_{A'}}^{W^{(-)};W{(+)}Z}\,
\left(1+\int_{a-i\infty}^{a+i\infty}
\frac{d\omega}{4i}
s^{\omega} \frac{1+e^{-i\pi \omega}}{\pi \omega}
\frac{\omega_c(q^2)}{\omega- \omega_c(q^2)} \right),
\label{reggeansatzcharge}
\end{equation}
where $a >0$. We can shift the integration contour to the region $\Re \omega <0$
by cancelling the result of taking the residue of the pole at $\omega =0$
with the Born contribution:
\begin{equation}
A_{11}(s,t)= \frac{2s}{-\bq^2 -M_W^2} 
\int_{-a-i\infty}^{-a+i\infty}
\frac{d\omega}{4i} \frac{1+e^{-i\pi \omega}}{\pi \omega}
s^{\omega}
g a_{\lambda_A}^{W^{(+)};W{(-)}Z} \frac{\omega _c(q^2)}{\omega- \omega_c(q^2)}
g a_{\lambda_{A'}}^{W^{(-)};W{(+)}Z}.
\label{deltaj}
\end{equation}
With the partial wave
\beq
F_{11}(\omega,q^2) = \frac{\omega _c(q^2)}{(\bq^2+M_W^2)(\omega- \omega_c(q^2))}
\eeq
we write the partial wave representation in the form:
\beq
A_{11}(s,t)= 2s
\int_{-a-i\infty}^{-a+i\infty}
\frac{d\omega}{4i} s^{\omega}\frac{1+e^{-i\pi \omega}}{\pi \omega}
g a_{\lambda_A}^{W^{(+)};W{(-)}Z} 
F_{11}(\omega,q^2)
g a_{\lambda_{A'}}^{W^{(-)};W{(+)}Z}.
\label{chargeSW}
\eeq
An analogous ansatz can also be made for production amplitudes. Note, that
the $t$-channel partial wave for the Born term
contains the Kronecker symbol $\sim \delta _{j,\,1}$ non-analytic in the
$j$-plane but as a result of summing radiative corrections the $t$-channel
partial wave in LLA becomes the analytic function \cite{GS79}
$$
\sim -\frac{\omega_c(q^2)}{\omega- \omega_c(q^2)}\,.
$$

From the point of view of the $t$-channel unitarity the
reggeization of the vector bosons is related to the existence of
the nonsense intermediate states for two  particles with spins
equal to unity \cite{GS79}. For these nonsense states the sum of
projections of their spins on the relative momentum
$\overrightarrow{p}$ equals 2, which makes them non-physical for
the total momentum $j=1$. Nevertheless, the $t$-channel partial
wave $f^{nn}_j(t)$ for the nonsence-nonsense transition exists for
complex $j$ and has the pole $\sim g^2/(j-1)$ in the Born
approximation. The $t$-channel unitarity condition together with
the dispersion relations allows one to construct this partial wave
in LLA: $f^{nn}_j(t)\sim g^2/(j-1-\omega (t))$, where $\omega (t)$
is the corresponding Regge trajectory. The similar calculation of
the amplitudes for  sense-nonsense and sense-sense transitions
gives in LLA $f_j^{ns}\sim g^2 \sqrt{j-1}/(j-1-\omega (t))$ and
$f_j^{ss}\sim \beta (t)/(j-1-\omega (t))$, respectively. It leads
to the disappearance of the singularity $\sim \delta _{j,\,1}$ in
the sense-sense partial wave \cite{GS79}.

In order to obtain the partial wave amplitude for neutral exchange in
the $2 \to 2$ process $ W^{(+)}W^{(-)} \to W^{(+)}W^{(-)}$, we return to  
(\ref{deltaj}).
Since, in the Born approximation, we have, instead of the propagator
$ \sim 1/(\bq^2 + M_W^2)$, the $Z$ and $\gamma$ propagators, we
replace the first term by $Z$ and $\gamma$
exchanges. Shifting then the $\omega$ contour to the left from
the point $\omega =0$, we arrive at
the form
\begin{align}
  \label{neutralSW}
  A_{10}(s,t) = 2 s\int _{-a-i\infty}^{-a+i\infty}\frac{d\omega}{4 i}
s^{\omega} \frac{1+e^{-i\pi \omega}}{\pi \omega}
 a_{\lambda_A}^{W_3;W^{(-)}W^{(+)}}
         F_{10}(\omega,q^2)
 a_{\lambda_A}^{W_3;W^{(+)}W^{(-)}}
             \nonumber \\
 + 2s\, \left(g a_{\lambda_A}^{Z;W^{(-)}W^{(+)}}\frac{c_w^2}{-\bq^2 -M_Z^2}
            g a_{\lambda_A}^{Z;W^{(+)}W^{(-)}}
+g a_{\lambda_A}^{\gamma;W^{(-)}W^{(+)}} \frac{s_w^2}{-\bq^2} g 
      a_{\lambda_A}^{\gamma;W^{(+)}W^{(-)}} \right. \nonumber\\ \left.
- g a_{\lambda_A}^{W_3;W^{(-)}W^{(+)}} \frac{1}{-\bq^2 - M_W^2}
g a_{\lambda_A}^{W_3;W^{(+)}W^{(-)}}\right)\,.
\end{align}
As a result of shifting the contour to the left we have obtained
from the residue of the pole $1/\omega$ the third contribution
$\sim 1/(\bq^2 + M_W^2)$ in the last brackets. All terms in these
brackets contain the non-analytic factors $\delta _{j,\,1}$ in the
$j$-plane (see the above discussion after eq. (\ref{deltaj})).
Thus, for $t<0$ the high energy behaviour of scattering amplitudes
$A\sim s$ for the neutral $t$-channel is governed by these
Kronecker-symbol singularities (see (\ref{neutralallorder})).

For the neutral exchange  the $t$-channel partial wave
in the Born approximation is not factorized \cite{GS79}. As a result,
the sense-sense amplitudes in LLA have both the Regge pole and the
Kronecker singularities.
The nonsense-nonsense partial wave for the neutral channel contains
the factor $(t-M^2)$, which leads
(after the use of the $t$-channel  unitarity condition) to the Regge trajectory
$\omega _n (q^2)$ proportional to this factor
(see (\ref{trajW3})) \cite{GS79}.

\subsection{Neutral isospin-1 channel}

Let us now turn to the integral equations.
We begin with the odd signature neutral exchange channel and 
consider the process $W^{(+)}W^{(-)} \to W^{(+)}W^{(-)}$.
The ansatz is contained in (\ref{neutralSW}). 
The $t$-channel partial wave $F$ is described by the sum of diagrams
illustrated in Fig.9a:
\beqn
  \label{neutralmellin}
  F_{10}(\omega,q^2)=\hspace{6cm}\nonumber \\ g^2 \sum_{n=0}^\infty \int
 \left( \prod_{l=1}^{n+1}
\frac{d^2q_{l}}{(2 \pi)^3} \frac{1}{(\bq_{l}^2+M_W^2)((\bq-\bq_{l})^2+M_W^2)
\left[ \omega-\omega_c(q_l^2)-\omega_c((q-q_l)^2) \right] } \right) g^2
\nonumber\\
\times K^{10}_{1,2}\cdot K^{10}_{2,3}\cdot ... \cdot K^{10}_{n,n+1}\,.
\hspace{4cm}
\eeqn
The kernel $K^{10}$ represents the sum of productions
of a $Z$ boson, a photon, and a Higgs scalar. It has the form:
\beqn
  \label{eq:neutralkern}
  K^{10}(q,k,k^\prime)&=&g^2 \Bigg\{ -(\bq^2+M_W^2)\\ \nonumber &+& \Big[
(\bk^2+M_W^2)((\bq-\bk^\prime)^2+M_W^2)+(\bk^{\prime
  2}+M_W^2)((\bq-\bk)^2+M_W^2)\Big]\\ \nonumber
&\times&
\left(\frac{c_w^2}{(\bk-\bk^\prime)^2+M_Z^2}+\frac{s_w^2}{(\bk-\bk^\prime)^2}
\right)
\Bigg\}\, .
\eeqn
It is convenient to remove, in (\ref{neutralmellin}), the first momentum
integral (in Fig.9a the leftmost cell), and to define the (amputated)
amplitude $f$:
\begin{equation}
  \label{neutralbeth}
  F_{10}(\omega,q^2)=\frac{g^2}{(2\pi)^3}\int
\frac{d^2k}{(\bk^{2}+M_W^2)((\bq-\bk)^2+M_W^2)}f_{10}(\omega;\bk,\bq-\bk)\,.
\end{equation}
For the amplitude $f_{10}(\omega;\bk,\bq-\bk)$ we can write down the
integral equation:
\begin{eqnarray}
  \label{eq:int10}
&&\hspace{-3.9cm}
[\omega-\omega_c(k^2)-\omega_c((q-k)^2)]f_{10}(\omega;\bk,\bq-\bk)= g^2
\\ \nonumber \\ \nonumber \hspace{2cm}
&+&\int \frac{d^2k^\prime}{(2\pi)^3}
\frac{K^{10}(\bq,\bk,\bk^\prime) \;
f_{10}(\omega;\bk^\prime,\bq-\bk^\prime)}
{(\bk^{\prime 2}+M_W^2)((\bq-\bk^\prime)^2+M_W^2)} \,.
\end{eqnarray}
The solution is independent of $k$, and we can easily find:
\beqn
  \label{eq:sol0}
  f_{10}(\omega,q^2)&=&\frac{g^2}{\omega-(q^2-M_W^2)\beta_{ww}(q^2)}
         \nonumber \\ &=&
\frac{g^2}{\omega-\omega_n(q^2)}\,,
\eeqn
and therefore
\beqn
F_{10}(\omega,q^2) = \frac{g^2 \beta_{ww}(g^2)}{\omega - \omega_n}\,.
\eeqn
This bootstrap solution reproduces exactly the Regge pole in the
neutral channel, which shows the self-consistency of our ansatz.

\subsection{Charged isospin-1 channel}

For the charged exchange channel we consider the process
$ZZ\to W^{(+)} W^{(-)}$ (and its $s - u$ counterpart $W^{(+)}Z\to ZW^{(+)}$).
The ansatz is contained in (\ref{chargeSW}).
The squared production amplitudes for the process $ZZ\to W^{(+)} W^{(-)}$
are illustrated in Fig.9b:
the left two kernels contain the production of a charged vector boson,
the next kernel contains the Higgs production. The partial wave has the
same structure as (\ref{neutralmellin}), but, as we said above, for each
neutral exchange we have to sum over three contributions: $Z$ and $\gamma$
exchanges in the Born approximation, and the neutral Regge exchange (in the
latter we have to subtract the particle pole).

To obtain an integral equation for the partial wave, in analogy to
(\ref{neutralbeth}), we remove the first (leftmost) loop integral;
in Fig.9b, the first cell has a charged $t$-channel
propagator below, a neutral one above. Denote the sum of the cells
to the right by $\tilde{f}_{cn_i}$ (here we include, 
for the coupling to the external particles on the rhs, 
the vertices from Table 1). For the crossed process, $W^{(-)}Z\to
ZW^{(-)}$, the first cell has the neutral $t$-channel propagator
below and the charged one above; let the sum of the cells to the
right be $\tilde{f}_{n_i c}$. In both cases, the subscript $i$ reminds
that, in the neutral exchange channel, we have to sum over several
terms ($Z$, $\gamma$, Regge pole minus particle pole): it will be
convenient to count the last term as a sum of two pieces. The
subscript $i$ then takes the four values: $i$ = $Z$, $\gamma$, $3$
(neutral particle pole), $n$ (neutral Regge pole). The last term
(neutral Regge pole) has the trajectory function
$\omega_{n}$, whereas for the other three terms the
trajectory is absent. It will be convenient to introduce,
nevertheless, the vanishing functions $\omega_Z = \omega_{\gamma}=\omega_3 =
0$. Also, each of the four terms has a multiplicative factor,
$b_i^2$: 
\beq b_Z^2 = c_w^2, \hspace{1cm}b_{\gamma}^2=s_w^2,
\hspace{1cm} b_3^2=-1, \hspace{1cm}b_n^2=1. \label{bconstants}
\eeq
With these considerations the coupled integral equations for
the functions $f_{n_i c}$ and $f_{c n_i}$ can be written in a
closed form:
\begin{equation*}
 [\omega -\omega_c(k^2)\;-\omega_{n_i}((q-k)^2)] \left(
\begin{array}{c} \tilde{f}_{cn_i} \\ \tilde{f}_{n_i c} \end{array} \right)
\end{equation*}
\begin{equation}
\label{eq:int11}
=\left( \begin{array}{c} g^2 a_ {\lambda_B}^{Z;HZ}  
a_{\lambda_B}^{W;WH}\delta_{iZ} \\
g^2a_{\lambda_B}^{W^{(-)};W^{+)}Z}a_{\lambda_B}^{n_i;W^{(+)}W^{(-)}}  \end{array} \right)
+ \left( \begin{array}{cc}
K_{cn_i;cn_j} & K_{cn_i;n_kc} \\ K_{n_ic;cn_j} & K_{n_ic;n_kc}
\end{array} \right) \otimes
\left( \begin{array}{c} \tilde{f}_{cn_j} \\ \tilde{f}_{n_k c} 
\end{array} \right)\,.
\end{equation}
The kernels $K_{cn_i;n_kc}$ etc. follow from the squares of production
vertices described before. The convolution symbol contains particle
propagators and the factors  $b_i^2$ (\ref{bconstants}).

Finally we define the signatured amplitudes: 
\beq
\tilde{f}_i^{(-)}(\bk,\bq -\bk ) = \tilde{f}_{c n_i}(\bk, \bq-\bk ) - 
                                 \tilde{f}_{n_i c}(\bq-\bk, \bk )\,.
\eeq 
These signatured partial waves satisfy the following integral
equations: 
\beq [\omega -\omega_c(k^2)\;-\omega_{n_i}((q-k)^2)]
\tilde{f}_i^{(-)}(\bk, \bq-\bk ) = g^2 a_{\lambda_B}^{W^{(-)};W^{(+)}Z}   +  \left( K_{ij} \otimes
\tilde{f}_j^{(-)}\right) (\bk,\bq-\bk)\,. 
\eeq 
We remove the vertex factor by rescaling the signatured amplitude and
obtain
\beq [\omega -\omega_c(k^2)\;-\omega_{n_i}((q-k)^2)]
{f}_i^{(-)}(\bk, \bq-\bk ) = g^2    +  \left( K_{ij} \otimes
{f}_j^{(-)}\right) (\bk,\bq-\bk)\,. 
\label{equationcharged} 
\eeq 
The kernels contain the sum of $W$ and Higgs production, and they are
of the form: 
\beqn K^{11}_{ij}(q,k,k^\prime)&=& g^2 \Bigg\{(-\bq^2
-M^2_W)\\ &+&
 \frac{(\bk^2+M_W^2)((\bq-\bk^\prime)^2+M_j^2)+(\bk^{\prime
  2}+M_W^2)((\bq-\bk)^2+M_i^2)}{(\bk-\bk^\prime)^2+M_W^2}  \Bigg\}
\nonumber
\eeqn
with $M_3=M_W$.

The bootstrap solution to this equation has the form:
\beq
\left(\begin{array}{c} f_Z^{(-)}(\bk,\bq-\bk)\\f_{\gamma}^{(-)}
(\bk,\bq-\bk)\\f_{3}^{(-)}(\bk,\bq-\bk)\\f_n^{(-)}(\bk,\bq-\bk)
\end{array} \right)
= \frac{g^2}{\omega - \omega_c(q^2)} \left(
 \left(\begin{array}{c} 1 \\1 \\ 1 \\ 1
 \end{array} \right)
 +\frac{\beta_{ww}((\bq-\bk)^2)}{(\omega-\omega_c(k^2))}
\left(\begin{array}{c} ((\bq-\bk)^2+M_Z^2) \\(\bq-\bk)^2  \\
         ((\bq-\bk)^2 + M_W^2) \\ 0  \end{array} \right)\,\right)\,.
\label{solutioncharged}
\eeq
When verifying that this solution satisfies the integral equation
(\ref{equationcharged}) it is useful to note the identities
\beq
b_Z^2 + b_{\gamma}^2 + b_3^2 = 0
\label{id1}
\eeq
and
\beq
b_Z^2 (\bq^2 + M_Z^2) + b_{\gamma}^2 \bq^2 + b_3^2 (\bq^2 + M_W^2) = 0.
\label{id2}
\eeq

With the solution (\ref{solutioncharged}) the partial wave becomes:
\beqn
F^{(-)}_{\omega;11}(q^2)&=& g^2 \int \frac{d^2 k}{(2 \pi)^3}
       \sum_i \frac{ b_i^2 f^{(-)}_i(\bk,\bq-\bk;\omega)}
{(\bk^2+M_W^2) ((\bq-\bk)^2+M_i^2)}\nonumber\\
&=& \frac{g^2 (c_w^2 \beta_{wz}(q^2)
+ s_w^2 \beta_{w\gamma}(q^2)) } {\omega - \omega_c(q^2)}\nonumber \\
& = &\frac{g^2 \omega_c(q^2)}{(-\bq^2 - M_W^2)(\omega -
\omega_c(q^2)}\,. 
\eeqn 
When going from the first to the
second line, we have used the identity (\ref{id1}).
This bootstrap relation 
completes our all-order proof of the reggeization for the weak
bosons.

\subsection{Vacuum channel}

Let us finally come to the zero quantum number exchange channel which
describes the 'electroweak Pomeron'. We consider, again, the process 
$W^{(+)}W^{(-)} \to W^{(+)}W^{(-)}$.
Since in the vacuum channel the signature is
positive, the Sommerfeld-Watson transform of the amplitude
at high energies reads:
\begin{eqnarray}
  \label{eq:watsonplus}
  A_{00}(s,t)=s\int\frac{d\omega}{2 i} \left(\frac s{M_W^2}\right)
  ^\omega \frac{-1+e^{-i\pi\omega}}{\pi \omega}
  F_{\omega,00}(q^2)\,.
\end{eqnarray}
The lowest-order diagrams to be summed are those of Fig.5e,f; in addition 
we have 
\begin{figure}
\begin{center}
\epsfig{file=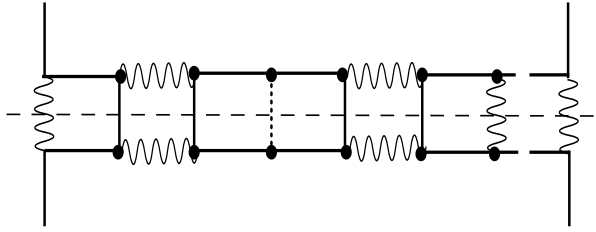,width=8cm,height=3cm}
\end{center}
\caption{Ladder diagrams obtained from the square of production amplitudes:
the vacuum exchange channel.}
\end{figure}
to include also $t$-channel contributions of two neutral exchanges (Fig.5g). 
The integral equation is illustrated in Fig.10. Using notations which are 
analogous to those of the previous subsection, we find 
the following set of integral equations:
\begin{align}
 \left( \begin{array}{l}
(\omega -\omega_{n_i}(k^2)\;-\omega_{n_j}((q-k)^2))\; f_{n_i n_j} \\ 
(\omega -\omega_c(k^2)\;-\omega_c((q-k)^2))\; f_{cc} \end{array}
\right) 
 =\left( \begin{array}{c}
g^2 a_{\lambda_B}^{n_i;W^{(+)}W^{(-)}} a_{\lambda_B}^{n_j;W^{(+)}W^{(-)}}
\\ - \frac{g^2}{\sqrt{2}} a_{\lambda_B}^{W_3; W^{(+)}W^{(-)}}
\end{array} \right) \nonumber \\
+ \left( \begin{array}{cc}
K^{00}_{n_in_j;n_{i'}n_{j'}} & \sqrt{2} K^{00}_{n_in_j;cc} \\
\sqrt{2} K^{00}_{cc;n_{i'}n_{j'}}
& K^{00}_{cc;cc} \end{array} \right) \otimes \left( \begin{array}{c}
f_{n_{i'}n_{j'}} \\ f_{cc}
\end{array} \right) .
\label{eq:vacuum}
\end{align}
Note that, in order to obtain these equations, one starts from a larger 
set of coupled equations: there are the two separate $t$-channels,
$W^{(+)}W^{(-)}$ and $W^{(-)}W^{(+)}$. Introducing even and odd combinations 
of them, the odd signature channel decouples, and one is left with 
(\ref{eq:vacuum}). These equations still contain all neutral even signature
channels, i.e. both the vacuum channel, $T=0$, $T_3=0$, and the
$T=2$, $T_3=0$ configurations. 
The kernels are derived from the production of vector bosons and of Higgs 
scalars, and they are easily obtained from our rules for production amplitudes.
Beginning with the kernel 
$K^{00}_{n_in_j;n_{i'}n_{j'}}$ which has neutral exchanges both on the 
left and on the right hand sides we note that these kernels are due
to the Higgs production only; whenever (at least) one of the exchanges
on the lhs or on the rhs is a photon, the matrix element vanishes. All other
kernels are found to have the structure 
\beq
K^{00}_{n_in_j;n_{i'}n_{j'}}= g^2\,\frac{M_W^2}{2c_w^{2m}}\,,
\label{K11} \eeq 
where $m$ is the total number of $Z$ lines (note
that the factor $1/2$ appears since we have included a factor $2$
into the integration phase space). As an example, 
\beq
K^{00}_{ZZ;ZZ}=g^2 \,\frac{M_W^2}{2c_w^8}, 
\eeq 
whereas 
\beq
K^{00}_{nn;ZZ}=g^2\,\frac{M_W^2}{2c_w^4}. 
\eeq

Next we list the kernels which have neutral exchanges on the lhs and 
charged exchange on the rhs,
$K^{00}_{n_in_j;cc}$. These matrix elements are symmetric under
the exchange of lhs and rhs: 
\beq
K^{00}_{n_in_j;cc} = K^{00}_{cc;n_in_j}\,. 
\eeq 
Their form follows
from (\ref{squarevertex1}), and the results can be summarized by:
$$
 K^{00}_{n_in_j;cc}(q,k,k^\prime)=
$$
\begin{equation}
g^2\left(- \bq^2 - M_{ij}^2+
 \frac{(\bk^2+ M_i^2)((\bq-\bk^\prime)^2+M_W^2)+(\bk^{\prime 2}+M_W^2)
  ((\bq-\bk)^2+M_j^2)}    {(\bk-\bk^\prime)^2+M_W^2}\right) \,,\nonumber
\label{K12}
\end{equation}
where
\begin{equation}
 M_{ij}^2=M_i^2+M_j^2-\frac{M_i^2M_j^2}{2M^2_W}\,.
\label{contact}
\end{equation}
In particular
\beqn
M_{ZZ}^2 &=& 2M_Z^2 - \frac{M_Z^4}{2M_W^2},\nonumber\\
M_{\gamma \gamma}^2 &=& 0, \nonumber\\
M_{nn}^2  &=& \frac{3}{2} M_W^2, \nonumber \\
M_{Z \gamma}^2 &=& M_Z^2,\nonumber \\
M_{Z n}^2 &=& M_W^2 + \frac{1}{2} M_Z^2, \nonumber\\
M_{\gamma n}^2 &=& M_W^2. 
\eeqn 
Finally, the kernel $K^{00}_{cc;cc}$ is the same as in 
(\ref{eq:neutralkern}), i.e.  
\beqn
  \label{eq:neutralkernrepeat}
   K^{00}_{cc;cc}(q,k,k^\prime)&=&g^2 \Bigg\{ - \bq^2 - M_W^2 \\ \nonumber &+& \Big[
(\bk^2+M_W^2)((\bq-\bk^\prime)^2+M_W^2)+(\bk^{\prime
  2}+M_W^2)((\bq-\bk)^2+M_W^2)\Big]\\ \nonumber
&\times&
\left(\frac{c_w^2}{(\bk-\bk^\prime)^2+M_Z^2}+\frac{s_w^2}{(\bk-\bk^\prime)^2}
\right)
\Bigg\}\, .
\eeqn

As we have said before, the integral equation (\ref{eq:vacuum}) still 
contains both the vacuum channel and the $T_3=0$ component of the 
$T=2$ channel. Nevertheless, we will refer to this set of equations as the 
electroweak 'vacuum exchange equation'. In order to separate the $T=0$ 
channel from the $T=2$ channel, we have to diagonalize the matrix equation. 
This cannot be done analytically, 
and in this paper we will discuss only a few approximations. 
First, in the case of vanishing Weinberg angle, $s_w=0, c_w=1$ in the
equation (\ref{eq:vacuum}), we can neglect the photon contributions
and leave only the reggeized boson $Z=W_3$ substituting
$n_i\rightarrow n,\,n_j \rightarrow n$. In this limit the kernel $K$ is
simplified as follows
\begin{equation}
\label{SU(2)}
K=
\left( \begin{array}{cc} K_{11} & \sqrt{2} K_{12} \\
\sqrt{2} K_{21} & K_{22}\end{array} \right),
\end{equation}
\newpage
\noindent
where
\begin{equation}
K_{11}=g^2\,\frac{M^2}{2}\,,\,\,\,
K_{12}=K_{21}=
\end{equation}
$$
g^2 \left(- \bq^2 - \frac{3}{2}M^2+
 \frac{(\bk^2+ M^2)((\bq-\bk^\prime)^2+M^2)+(\bk^{\prime 2}+M^2)
  ((\bq-\bk)^2+M^2)}    {(\bk-\bk^\prime)^2+M^2}\right)\,,
$$
$$
K_{22}=g^2\,\left(- \bq^2 - M^2+
 \frac{(\bk^2+ M^2)((\bq-\bk^\prime)^2+M^2)+(\bk^{\prime 2}+M^2)
  ((\bq-\bk)^2+M^2)}    {(\bk-\bk^\prime)^2+M^2}\right)\,.
$$
This kernel is $SU(2)$-invariant, and we can search the solution
of the corresponding equation for the vacuum exchange in the form
\begin{equation}
\left( \begin{array}{c}
f_{n n} \\
f_{cc} \end{array} \right)=\left( \begin{array}{c}
1 \\
\sqrt{2} \end{array} \right)\,f_{WW}\,.
\label{ansatzT=0}
\end{equation}
For the function $f_{WW}$ we obtain the known BFKL equation
for the Pomeron wave function in the $SU(2)$ case \cite{BFKL2}
$$
(\omega -\omega (k^2)\;-\omega ((q-k)^2)) f_{WW}(k,q)=
$$
\begin{equation}
g^2 a_{\lambda_B}^{0;W^{(+)}W^{(-)}} +
2 \int \frac{d^2k'}{(2\pi)^3} K_{BFKL}
\frac{1}{(\bk^\prime )^2+M^2}\frac{1}{(\bq-\bk^\prime )^2+M^2}
f_{WW}(k^\prime ,q)\,,
\label{SU2BFKL}
\end{equation} 
where the integral kernel is given by:
$$ K_{BFKL} = $$
\beq
g^2\,\left(- \bq^2 - \frac{5}{4}M^2+
 \frac{(\bk^2+ M^2)((\bq-\bk^\prime)^2+M^2)+(\bk^{\prime 2}+M^2)
  ((\bq-\bk)^2+M^2)}    {(\bk-\bk^\prime)^2+M^2}\right)\,,
\label{kernT0}
\eeq
and the couplings $a_{\lambda_B}^{0;W^{(+)}W^{(-)}}$ have the values 
$\frac{2}{\sqrt{3}}$, $\frac{3}{4\sqrt{3}}$ for $\lambda_B =1,2$ 
and $\lambda_B =3$, resp.
In (\ref{SU2BFKL}), the factor $2$ in front of the integral corresponds to
the $SU(2)$ group factor (cf. $r^{(0)}=2$ in (\ref{SU2decomp})).

The second solution to the matrix equation (\ref{SU(2)}) belongs to
the $T_3=0$ component of the $T=2$ representation. The corresponding
eigenvector, in analogy with (\ref{ansatzT=0}), is of the form:
\beqn
\left( \begin{array}{c}
f_{n n;T=2} \\
f_{cc; T=2} \end{array} \right)=
\left( \begin{array}{c} - \sqrt{2} \\ 1 \end{array} \right) \,f_{WW;T=2}\,.
\eeqn
The integral equation has the same form as (\ref{SU2BFKL}), where the
$SU(2)$ group factor in front of the integral is $-1$, and in the expression
(\ref{kernT0}) for
the kernel, the mass term $-\frac{5}{4}M^2$ is replaced by
$-2M^2$. 

As a second approximation, let us return to the investigation of  
the realistic case of the electroweak theory, eqs.(\ref{eq:vacuum}),
and find a somewhat simpler form.
To begin with, we note, that the inhomogeneous term does not depend
on the momenta $k$ and $q-k$ and corresponds to a local interaction
of the vector particles. Similarly, in the kernel $K$ the
matrix element $K^{00}_{n_in_j;n_{i'}n_{j'}}$ (\ref{K11}) and the
contribution proportional to $M_{ij}^2$ in expression (\ref{K12})
describe their contact interaction. We can take into account this
contributions to the kernel later restricting ourselves initially
to the solution of the equation, in which the kernel $\widetilde{K}$
does not contain these terms:
$$
\widetilde{K}_{11}=0\,,\,\,\,
\widetilde{K}_{12}=(\widetilde{K}_{21})_{k\leftrightarrow k'}=
$$
$$
g^2\sqrt{2}\left(- \bq^2 +
 \frac{(\bk^2+ M^2_i)((\bq-\bk^\prime)^2+M^2_W)+(\bk^{\prime 2}+M^2_W)
  ((\bq-\bk)^2+M^2_j)}{(\bk-\bk^\prime)^2+M^2_W}\right)\,,
$$
$$
\widetilde{K}_{22}=K^{10}(q,k,k')\,,
$$
where $K^{10}(q,k,k')$ is given in (\ref{eq:neutralkern}).
It is convenient also to write the equation (\ref{eq:vacuum}) as follows
\begin{equation}
\label{eq:vacsimp}
\left( \begin{array}{c}
\phi _{n_i n_j} \\
\phi _{cc} \end{array} \right)
 =\left( \begin{array}{c}
g^2\\ g^2 \end{array} \right) + \left( \begin{array}{cc}
\widetilde{K}_{11}& \widetilde{K}_{12}\\
\widetilde{K}_{21}
& \widetilde{K}_{22} \end{array} \right) \otimes \left( \begin{array}{c}
\phi _{n_{i'}n_{j'}}/(\omega -\omega _{i'}-\omega _{j'}) \\
\phi _{cc}/(\omega -\omega _c-\omega _c)
\end{array} \right),
\end{equation}
where the reggeon propagators in the right hand side of the equation are 
integrated over $k'$ (for simplicity, we here disregard the helicity structure 
of the couplings to external particles). Looking at
the kernel $\widetilde{K}$, we see, that only the matrix element
$\widetilde{K}_{12}$ depends on $i$ and $j$, and this dependence
is simple. It means, that we can search a solution of the above equation
in the form
\begin{equation}
\label{anzatz}
\left( \begin{array}{c}
\phi _{n_i n_j} (k,q)\\
\phi _{cc} (k,q)\end{array} \right)
 =\left( \begin{array}{c}
\phi _0(k,q)+M_i^2\,\phi _1(k,q)+M_j^2\,\phi _2(k,q)\\ \phi _{cc}
\end{array} \right) ,
\end{equation}
where
\begin{equation}
\phi _2(k,q)=\phi _1(q-k,q)\,.
\end{equation}
Putting this ansatz in the equation, we obtain the system of the
equations for the functions $\phi _i$:
\begin{equation}
\phi _0(k,q)=g^2\sqrt{2}\left(- \bq^2 +
 \frac{\bk^2((\bq-\bk^\prime)^2+M^2_W)+(\bk^{\prime 2}+M^2_W)
  (\bq-\bk)^2}{(\bk-\bk^\prime)^2+M^2_W}\right)\otimes
\frac{\phi _{cc}}{\omega -\omega _c -\omega _c}\,,
\end{equation}
\begin{equation}
\phi _1(k,q)=g^2\sqrt{2}
\frac{(\bq-\bk^\prime)^2+M^2_W)}{(\bk-\bk^\prime)^2+M^2_W)}\,
\otimes  \frac{\phi _{cc}}{\omega -\omega _c -\omega _c}
\end{equation}
\begin{equation}
\phi _{cc}(k,q)=g^2+\widetilde{K}_{21} \otimes
\frac{\phi _{n_{i'}n_{j'}}}{\omega -\omega _{i'} -\omega _{j'}}+
\widetilde{K}_{22} \otimes
\frac{\phi _{cc}}{\omega -\omega _c -\omega _c}\,.
\end{equation}
Here the integration over $k'$ is implied. Returning to the
general case (\ref{eq:vacuum}), we note, that the inhomogeneous
terms and the terms $K^{00}_{n_in_j;n_{i'}n_{j'}},\,M_{ij}^2$ in
the kernel lead to the diagrams, in which the ladders generated by
the simplified kernel $K$ are combined each with other by the
local vertices. Therefore the solution of (\ref{eq:vacuum}) can be
obtained from the solutions of the simplified equation
(\ref{eq:vacsimp}) with modified inhomogeneous terms by summing
the corresponding two-point loop diagrams. We consider this
procedure in details in our future publications.

Finally we note that the equations (\ref{eq:vacuum}) simplify in
the region of large transverse momenta \beq \bk^2,\; (\bq-\bk)^2\;
\gg M_i^2\,, \eeq where we can neglect all masses and Higgs
contributions. In this region, in each neutral exchange channel
the sum of the non-reggeizing pieces cancels (due to (\ref{id1})),
and we are left with the Regge pole $\omega_n$ only. Consequently
we are back to the massless $SU(2)$ gauge theory, i.e. the kernels
are conformal invariant. The leading high energy behaviour in the
vacuum channel then follows from the observation that, because of
the diffusion in $\ln \bk^2$, the spectrum of eigenvalues is the
same as in the massless case:
\begin{equation}
\omega =\omega (\nu ,n)=\frac{g^{2}}{\pi ^{2}}\left( \Psi (1)-Re \,\Psi (%
\frac{1}{2}+i\nu +\frac{\left| n\right| }{2})\right) \,,
\end{equation}
where $\nu $ and $n$ are real and integer numbers, resp. The leading
singularity of the $t$-channel partial wave appears at
\begin{equation}
\omega =\omega (0,0)=2\,\frac{g^{2}}{\pi ^{2}}\,\ln \,2\,
\end{equation}
and leads to the power-like behaviour $\sigma _{t}\sim s^{\omega
(0,0)}$ of the total cross-sections. Note, however, that the
solution of equation (\ref{eq:vacuum}) can contain the Regge poles
at $\omega =\omega _0 > \omega (0,0)$ with the residues tending to
zero at $k^2 \rightarrow \infty$. Further investigations of the
related problems are in progress, including a numerical solution
to the coupled integral equations.

\section{An application: WW-scattering}

At the end of our paper we present, as an application of the vacuum 
channel integral equation, the two loop expressions for 
the process $W^{(+)}+W^{(-)}  \to W^{(+)}+W^{(-)}$.
This elastic scattering process has, as 'secondary Regge' exchange, 
the odd signature neutral isospin-$1$ exchange, described in section 5.1. 
For the even signature part the combined $T=0$ and $T=2$ exchanges, in the 
one-loop approximation, are given in (\ref{vacuum1loop}). 
The higher-loop approximations can be derived from (\ref{eq:vacuum}) 
which we rewrite in the following way:
\begin{align}
 \omega \left( \begin{array}{l}
f_{n_i n_j} \\ 
f_{cc} \end{array}
\right) 
 =\left( \begin{array}{c}
g^2 a_{\lambda_B}^{n_i;W^{(+)}W^{(-)}} a_{\lambda_B}^{n_j;W^{(+)}W^{(-)}}
\\ - \frac{g^2}{\sqrt{2}} a_{\lambda_B}^{W_3; W^{(+)}W^{(-)}}
\end{array} \right) \nonumber \\
+ \left( \begin{array}{cc}
K^{00}_{n_in_j;n_{i'}n_{j'}} + (\omega_{n_i} +\omega_{n_j}) \delta_{ii'}
\delta_{jj'} & \sqrt{2} K^{00}_{n_in_j;cc} \\
\sqrt{2} K^{00}_{cc;n_{i'}n_{j'}}
& K^{00}_{cc;cc} +\omega_c +\omega_c \end{array} \right) \otimes \left( \begin{array}{c}
f_{n_{i'}n_{j'}} \\ f_{cc}
\end{array} \right)\,.
\label{eq:vacuumprime}
\end{align}
For the couplings to external particles 
we introduce column vectors ('impact vectors'), 
$\Phi_{W^{(+)}}$ and $\Phi_{W^{(-)}}$: 
for the $W^{(-)}$, $\Phi_{W^{(-)}}$ is given by the inhomogenous term on the 
rhs of (\ref{eq:vacuumprime}), whereas for the $W^{(+)}$ 
we have 
\beqn
\Phi_{W^{(+)}}=
\left( \begin{array}{c}
g^2 a_{\lambda_A}^{n_i;W^{(-)}W^{(+)}} a_{\lambda_A}^{n_j;W^{(-)}W^{(+)}}
\\ \frac{g^2}{\sqrt{2}} a_{\lambda_A}^{W_3; W^{(-)}W^{(+)}}
\end{array} \right)\,.
\eeqn 
The two-loop approximation, in a symbolic notation, is then simply
given by 
\beqn  
A_{even}^{(2)}\;=\;
\Phi_{W^{(+)}}^T \otimes K \otimes \Phi_{W^{(-)}}\,,
\eeqn
where $K$ denotes the matrix kernel of (\ref{eq:vacuumprime}).
After some algebra we find: 
\beqn
A_{even}^{(2)}\;=\; 2 i \pi s \left( (a_A^Z)^2\beta_{ZZ} M_W^2 \beta_{ZZ}
(a_B^Z)^2 
- \frac{1}{2} a_A^{W_3} \beta_{WW} (-\bq^2 -M_W^2) \beta_{WW} a_B^{W_3} 
\hspace{4cm}
\right.\nonumber\\ \left.
+ \frac{1}{\sqrt{2}} a_A^{W_3} \beta_{WW} [(-\bq^2 - M_{ZZ}^2) c_w^4
\beta_{ZZ} (a_B^{Z})^2 +
(-\bq^2 - M_{\gamma\gamma}^2) s_w^4 \beta_{\gamma\gamma}(a_B^{\gamma})^2
\hspace{3.5cm}
\right.\nonumber\\ \left.          
+2(-\bq^2 - M_{\gamma Z}^2) \beta_{\gamma Z} c_w^2 s_w^2 
          a_B^{Z}a_B^{\gamma}] 
\hspace{5.5cm}
\right.\\ \left.    
- \frac{1}{\sqrt{2}}   [(a_A^Z)^2 c_w^4 \beta_{ZZ} (-\bq^2 - M_{ZZ}^2) 
 + (a_A^{\gamma})^2    s_w^4 \beta_{\gamma\gamma} (-\bq^2 - M_{\gamma\gamma}^2)
\hspace{5cm}
\right.\nonumber\\ \left.          
+2 a_A^Z a_A^{\gamma}
c_w^2 s_w^2 \beta_{\gamma Z} (-\bq^2 - M_{\gamma Z}^2)] \beta_{WW} a_B^{W_3}
\hspace{4cm}
\right.\nonumber\\ \left.
+ 2 \beta_{WWZ} c_w^2 [- a_A^{W_3} a_A^Z a_B^{W_3} a_B^Z
+\frac{1}{\sqrt{2}} a_A^{W_3} (c_w^2 (a_B^{Z})^2 + s_w^2 a_B^{Z} a_B^{\gamma}) 
- \frac{1}{\sqrt{2}} (c_w^2 (a_A^{Z})^2 +s_w^2 a_A^{Z} a_A^{\gamma} ) 
a_B^{W_3} ]
\right.\nonumber\\ \left.  
+ 2 \beta_{WW\gamma} s_w^2 [-_A^{W_3} a_A^{\gamma} a_B^{W_3} a_B^{\gamma} 
+\frac{1}{\sqrt{2}} a_A^{W_3} (s_w^2 (a_B^{\gamma})^2 + c_w^2 a_B^{\gamma}  
 a_B^{Z})
- \frac{1}{\sqrt{2}} (s_w^2 (a_A^{\gamma})^2 + c_w^2
a_A^{\gamma} a_A^{Z}) a_B^{W_3} ] \right)\,. \nonumber
\eeqn  
Here we have used the abbreviations $a_{\lambda_A}^{Z;W^{(-)} W^{(+)}} 
\to a_A^Z$ , $a_{\lambda_B}^{Z;W^{(+)} W^{(-)}} 
\to a_B^Z$ etc.
\section{Conclusions}
In this paper we have examined, in the leading logarithmic approximation,
the high energy behavior of the
electroweak sector of the Standard Model. We have derived bootstrap equations 
which describe the reggeization of the vector
bosons. The charged $W$ bosons
lie on the Regge trajectory $\alpha_c(t)$ which at $t=M_W^2$ passes through
unity. In the neutral
sector there exists another Regge trajectory, $\alpha_{n}$, which also
at $t=M_W^2$ passes through $1$, but neither the $Z$ boson nor the photon
lie on this trajectory. For finite $t$ both trajectories differ
from each other, thus reflecting the breaking of the gauge symmetry
$SU(2) \times U(1)$. As usual, the Reggeization of the electroweak
gauge bosons hints at some form of compositeness.
Note, that in the Grand Unified Theories all particles,
including the $Z$-boson and photon, lie on their Regge trajectories
\cite{GS79}.

Our main result is the integral equation for the even signature 
exchange in electroweak theory, which contains both the Pomeranchuk 
singularity and the zero component of the $T=2$ exchange.
One of the features of this equation is that,
in the region of large transverse momenta,
a conformal structure emerges, analogous to the one
of the QCD BFKL Pomeron. This suggests that, in the combined limit of high
energies and small distances, not only the strong sector but also the
electroweak sector of the Standard Model exhibits a deeper symmetry
pattern which could be related to string theory.\\ \\
{\bf Acknowledgements:} One of us (L.N.L.) thanks the
Alexander von Humboldt-Foundation for financial support, and the
II.Institut of Theoretical Physics, University Hamburg, and DESY for the
hospitality.\\ \\
\renewcommand{\theequation}{A\arabic{equation}}
\setcounter{equation}{0}
\section*{Appendix}
In this appendix we list a few details of the calculation of the one
loop corrections to $2\to3$ production amplitudes. The production amplitude
in the Born approximation has the factorized form (\ref{2tonborn}),
and we want to compute the two-particle
intermediate state unitarity integral in $(23)$ subsystem (Fig.11).
\begin{figure}
\begin{center}
\epsfig{file=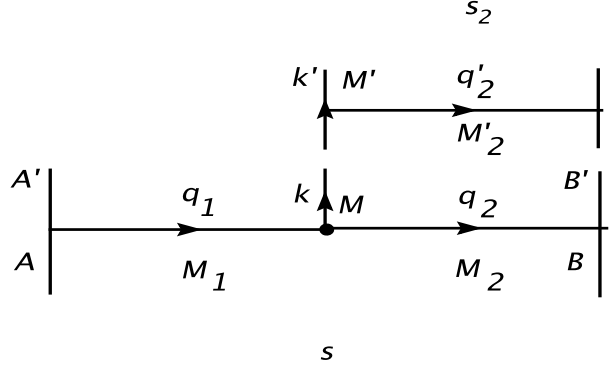,width=8cm,height=5cm}
\end{center}
\caption{Notations for the unitarity integral in the $(23)$-subchannel.
The dot denotes the effective production vertex.}
\end{figure}
Since the
the production amplitude (\ref{2tonborn}) holds in the overall cm-system,
whereas the $2\to2$ scattering amplitude (\ref{2to2born}) refers to the
cm-system of the $(23)$ subchannel,
it is necessary to transform from one reference frame to the other.
Following the discussion of ~\cite{Ba79}, we first compute the helicity
matrix elements
of the effective production vertex (\ref{effvertex}). We define the
polarization vectors
\beqn
e_{\mu}^1(k) &=&\frac{1}{|k|}\left( 0,\frac{k_1k_3}{|\bk|},
             \frac{k_2k_3}{|\bk|}, -|\bk|\right)\,, \nonumber\\
e_{\mu}^2(k) &=& \frac{1}{|\bk|} \left( 0,-k_2,k_1,0 \right)\,, \nonumber \\
e_{\mu}^3(k) &=& \frac{1}{M} \left( |k|, \frac{k_0 k_1}{|k|}\,,
\frac{k_0 k_2}{|k|}, \frac{k_0 k_3}{|k|} \right)\,,
\eeqn
where
\beq
k_0 = \frac{s_{1} + s_{2}}{2\sqrt{s}},\;\;
k_3 = \frac{s_{2} - s_{1}}{2\sqrt{s}}\,.
\eeq
The three helicity components of the production vertex are:
\beqn
C^{M;M_2M_1}(q_2,q_1)\cdot e^1(k) &=&
\frac{|\bk|}{2\sqrt{s}|k|}\left( s_{2} (1 +2\frac{t_1-M_1^2}{\bk^2+M^2})
                             + s_{1}(1 +2\frac{t_2-M_2^2}{\bk^2+M^2}) \right.
\,, \nonumber \\
             & & +\left. (s_{2}-s_{1}) \frac{t_2 -t_1}{\bk^2} \right)
\,,\nonumber \\
C^{M;M_2M_1}(q_2,q_1)\cdot e^2(k) &=& - 2 |\bq_1| \sin (\bq_1,\bk)
\nonumber \\
C^{M;M_2M_1}(q_2,q_1)\cdot e^3(k) &=& \frac{M }{2\sqrt{s}|k|}
\left( - s_{2} (1 +2\frac{t_1-M_1^2}{\bk^2+M^2})
                             + s_{1}(1 +2\frac{t_2-M_2^2}{\bk^2+M^2}) \right.
\nonumber \\
             & & +\left. (s_{2}+s_{1}) \frac{M_2^2 -M_1^2}{M^2} \right)
\eeqn
with
\beq
\bk^2 +M^2 =\frac{s_{1}s_{2}}{s}.
\eeq
An explicit calculation shows that the $2 \to 2$ subprocess in the $s_2$
channel, evaluated as a $3\times 3$ matrix in the overall cm-system,
can be written in the form
\beqn
L_{23} R(\bk,\bq'_2)  \left( \begin{array}{ccc} -1&0&0\\
                                                0&-1&0\\
                        0&0& -\frac{M^2+{M'}^2}{2MM'} + \frac{{M_2}^2}{2MM'}
                  \end{array} \right)
R^T (\bk',\bq'_2) {L_{23}'}^T\,,
\label{Ltrans}
\eeqn
where
\beq
L_{23}=\left( \begin{array}{ccc}
            A & 0 & -B \\
            0 & 1 &  0 \\
            B & 0 & A   \end{array} \right)
\label{Lmatrix}
\eeq
with
\beq
A= \frac{\sqrt{s}}{s_{2}|k|} \left( - M^2 +
          \frac{s_{2}}{2s} (s_{1} + s_{2})\right),\,\, B = \frac{\sqrt{s} |\bk| M}
{s_{2} |k|}.
\label{Lmatrixelements}
\eeq
The matrix $L'_{23}$ is obtained from $L_{23}$ by replacing $s_1 \to s'_1$
(with $\frac{s'_1 s_2}{s} = {\bk'}^2 + {M'}^2$). One easily verifies
unitarity, $LL^T=1$. The matrix $R(\bk, \bq'_2)$
denotes a rotation in the subspace of the transverse helicities:
\beq
R(\bk,\bq'_2) = \left( \begin{array}{ccc}
            \cos(\bk,\bq'_2) & \sin(\bk,\bq'_2)  & 0 \\
            - \sin(\bk,\bq'_2) & \cos(\bk,\bq'_2)   & 0 \\
            0 & 0 & 1   \end{array} \right).
\eeq
On the rhs of (\ref{Ltrans}), the matrix in the middle represents the
$2 \to 2$ scattering in the $s_2$ cm-system.
>From this result we infer that the Lorentz transformation,
which takes us from the $s_2$ cm-system of the
outgoing particles $2$ and $3$ into the overall cm-system of
the two incoming particles, consists of a rotation and of a boost.
Since in (\ref{Ltrans}) the two rotations commute with the
scattering matrix in the $s_2$ system, they can be combined into
\beq
R(\bk,\bq'_2) R^T (\bk',\bq'_2) = R(\bk,\bk'),
\eeq
and, in (\ref{Ltrans}), this matrix be written either on the lhs or on the rhs
of the diagonal $2 \to 2$ scattering matrix.
We also find that, in the double-Regge limit, the particle-reggeon-particle
vertex at the rhs of Fig.11 does not change if we switch from the overall
cm-system to the $s_2$ cm-system.

Multiplying now the vector of helicity matrix elements by
$L_{23}^{-1} = L_{23}^{T}$,
we obtain for the effective production vertex in the
$(23)$ cm-system:
\beqn
L_{23}^T \left( \begin{array}{c} C e^1\\ C e^2\\ C e^3 \end{array} \right)
=
 \left( \begin{array}{c} 2 |\bq_1| \bV(\bq_1,\bk) \\
M\, ( -1+ \frac{M_2^2 - M_1^2}{M^2})  \end{array} \right)
- \frac{\bq_1^2 +M_1^2}{\bk^2 + M^2}
\left( \begin{array}{c} 2 |\bk| \bV(\bk,\bk)  \\  -2M \end{array} \right)\,,
\label{vertex23cmsystem}
\eeqn
where
\beqn
\bV(\bq_1,\bk) =
\left( \begin{array}{c} \cos (\bq_1,\bk)\\
- \sin(\bq_1,\bk) \end{array} \right)\,.
\eeqn

Now we are ready to multiply with the $2 \to 2$ matrix element in the
$(23)$ channel and to compute the unitarity integral, using,
in particular, for the longitudinal component, the helicity factors of
Table 1. We first
consider the processes shown in Fig.7 where all wavy lines stand for $Z$
bosons. In the notation of this appendix we have:
$M_1\,=\,M\,=\,M'_2\,=\,M_W$, $M_2\,=\,M'\,=\,M_Z$, i.e. we start from
the effective vertex $C^{M_W M_Z M_W}$.
On the rhs of (\ref{vertex23cmsystem}), the third component of the first vector
becomes
\beq
M_W (-1 + \frac{M_Z^2 - M_W^2}{M_W^2} ) = 2M_W a_3^{Z;W^{(+)}W^{(-)}}
= 2 M_Z c_w a_3^{Z;W^{(+)}W^{(-)}}\,.
\eeq
Multiplying with the $W^{(+)}W^{(-)}\to ZZ$ matrix element and including the 
Higgs intermediate state, we obtain, at the production vertex:
\beqn
\left( \begin{array}{c} 2 |\bq_1| a_1 ^{W^{(-)};ZW^{(+)}} \bV(\bq_1,\bk) \\
2 M_W (a_3^{Z;W^{(+)}W^{(-)}} a_3^{W^{(-)};ZW^{(+)}} + a_3^{W;HW} a_3^{Z;HZ})   \end{array} \right)
- \frac{\bq_1^2 +M_W^2}{\bk^2 + M_W^2}
\left( \begin{array}{c} 2 |\bk| a_1^{W^{(-)};ZW^{(+)}} \bV(\bk,\bk)  
\\  -2M_W a_3^{W^{(-)};ZW^{(+)}} \end{array} \right)\,,
\eeqn
which can also be written as
\beqn
- c_w \Bigg[ \left( \begin{array}{c} 2 |\bq_1| \bV(\bq_1,\bk) \\
- M_Z   \end{array} \right)
- \frac{\bq_1^2 +M_W^2}{\bk^2 + M_W^2}
\left( \begin{array}{c} 2 |\bk| \bV(\bk,\bk)  \\  - M_Z
\end{array} \right) \Bigg]\,.
\eeqn
By subtracting and adding a new term, we arrive at the expression:
\beqn
- c_w \Bigg[ \left( \begin{array}{c} 2 |\bq_1| \bV(\bq_1,\bk) \\
- M_Z   \end{array} \right)
- \frac{\bq_1^2 +M_W^2}{{\bk'}^2 + M_Z^2}
\left( \begin{array}{c} 2 |\bk'| \bV(\bk,\bk')  \\  - 2M_Z
\end{array} \right) \Bigg] \nonumber \\
+ c_w  \frac{\bq_1^2 +M_W^2}{\bk^2 + M_W^2}
\Bigg[ \left( \begin{array}{c} 2 |\bk| \bV(\bk,\bk) \\
- M_Z   \end{array} \right)
- \frac{\bk^2 +M_W^2}{{\bk'}^2 + M_Z^2}
\left( \begin{array}{c} 2 |\bk'| \bV(\bk,\bk')  \\  - 2M_Z
\end{array} \right) \Bigg]\,.
\eeqn
The overall factor $-c_w= T^Z_{W^{(-)}W^{(+)}}c_w$ contains, apart from the 
isospin group factor, the wave function of the produced $Z$ boson, $c_w$.
The expression in the first and in the second lines can then be recognized 
as the result of the Lorentz transform applied to the effective production 
vertex $C^{M_Z; M_W M_W}$,

Combining with the other parts of Fig.11 and adding the corresponding
expression for the photon exchange  we find the following results
for the discontinuities in $s_{2}$  (Fig.12):
\begin{figure}
\begin{center}
\epsfig{file=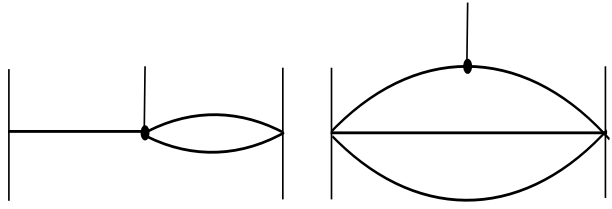,width=8cm,height=3cm}
\caption{Transverse momentum structure of eq.(A16)}
\end{center}
\end{figure}
\beq
disc_{s_2} A_{2 \to 3} = 2\pi s \frac{F_L}{t_1-M_1^2}
\eeq
with
\beqn 
F_L= -\frac{1}{2} s g a_{\lambda_A}^{W^{(-)};W^{(+)}Z} \left(
          g T^{Z}_{W^{(-)}W^{(+)}} C (\bq_2,\bq_1)^{M_Z;M_W M_W} 
\frac{\omega_{c}(q_2^2)}
              {-\bq_2^2 -M_W^2} \right.\nonumber \\
\left.
- (\bq_1^2 + M_W^2) (c_w^2K^{Z;WW} +s_w^2 
K^{\gamma;WW}) \right)
   g a_{\lambda_B}^{W^{(+)};W^{(-)}Z}
\eeqn
and
\beq
K^{Z;WW} = g^2 T^{Z}_{W^{(-)}W^{(+)}} \int \frac{d^2 k}{(2\pi)^3}
\frac{C^{M_Z; M_W M_W}(\bq_2-\bk,\bq_1-\bk)}
                     { ((\bq_1-\bk)^2 + M_W^2) ((\bq_2-\bk)^2 + M_W^2)}
                    \frac{1}{(\bk^2 +M_Z^2)}\,.
\eeq
The same calculations can be done for the other discontinuities illustrated 
in Figs.7 and 8. They lead to the results listed in section 4.


\begin{thebibliography}{99}
\bibitem{GST73} M.\thinspace T.~Grisaru, H.\thinspace J.~Schnitzer,%
H.\thinspace S.~Tsao, {\it Phys. Rev. Lett.}.\ {\bf 30}, 811 (1973).
\bibitem{Lip76}  L.\thinspace N.~Lipatov, {\it Sov. J. Nucl. Phys}.\ {\bf 23}%
, 338 (1976).
\bibitem{BFKL2}
V.\thinspace S.~Fadin, E.\thinspace A.~Kuraev and L.\thinspace
N.~Lipatov, {\it Phys. Lett}.\ B {\bf 60}, 50 (1975);
{\it Sov. Phys. JETP} .\ {\bf 44}, 443 (1976);
{\it Sov. Phys. JETP} .\ {\bf 45}, 99 (1977);
\bibitem{BL}
I.\thinspace
I.~Balitsky and L.\thinspace N.~Lipatov, {\it Sov. J. Nucl. Phys}.\ {\bf 28}%
, 822 (1978); {\it JETP Lett}.\ {\bf 30}, 355 (1979).
\bibitem{GS79} M.\thinspace T.~Grisaru, H.\thinspace J.~Schnitzer,
{\it Phys.Rev. } {\bf D 20}, 784 (1979).
\bibitem{LS} L.Lukaszuk and L.Szymanowski,
{\it Nucl.Phys.} {\bf B 159}, 316 (1979).
\bibitem{Ba79} J.Bartels, {\it Nucl.Phys.} {\bf B 151}, 293 (1979).
\end{thebibliography}
\end{document}